# Detailed cool star flare morphology with CHEOPS and TESS⋆

G. Bruno[1], I. Pagano[1], G. Scandariato[1], H.-G. Florén[2], A. Brandeker[2], G. Olofsson[2], P. F. L. Maxted[3], A. Fortier[4,5], S. G. Sousa[6], S. Sulis[7], V. Van Grootel[8], Z. Garai[9,10,11], A. Boldog[12,13,9], L. Kriskovics[12,13], Gy. M. Szabó[10,9], D. Gandolfi[14], Y. Alibert[5,4], R. Alonso[15,16], T. Bárczy[17], D. Barrado Navascues[18], S. C. C. Barros[6,19], W. Baumjohann[20], M. Beck[21], T. Beck[4], W. Benz[4,5], N. Billot[21], L. Borsato[22], C. Broeg[4,5], A. Collier Cameron[23], Sz. Csizmadia[24], P. E. Cubillos[25,20], M. B. Davies[26], M. Deleuil[7], A. Deline[21], L. Delrez[27,8,28], O. D. S. Demangeon[6,19], B.-O. Demory[5,4], D. Ehrenreich[21,29], A. Erikson[24], J. Farinato[22], L. Fossati[20], M. Fridlund[30,31], M. Gillon[27], M. Güdel[32], M. N. Günther[33], A. Heitzmann[21], Ch. Helling[20,34], S. Hoyer[7], K. G. Isaak[33], L. L. Kiss[35,36], K. W. F. Lam[24], J. Laskar[37], A. Lecavelier des Etangs[38], M. Lendl[21], D. Magrin[22], C. Mordasini[4,5], V. Nascimbeni[22], R. Ottensamer[32], E. Pallé[15,16], G. Peter[39], G. Piotto[22,40], D. Pollacco[41], D. Queloz[42,43], R. Ragazzoni[22,40], N. Rando[33], F. Ratti[33], H. Rauer[24,44], I. Ribas[45,46], N. C. Santos[6,19], M. Sarajlic[3], D. Ségransan[21], A. E. Simon[4,5], V. Singh[1], A. M. S. Smith[24], M. Stalport[8,27], N. Thomas[4], S. Udry[21], B. Ulmer[39], J. Venturini[21], E. Villaver[15,16], N. A. Walton[47], T. G. Wilson[41]

(Affiliations can be found after the references)

Received xxx; accepted yyy

**ABSTRACT**

*Context.* White-light stellar flares are proxies for some of the most energetic types of flares, but their triggering mechanism is still poorly understood. As they are associated with strong X and ultraviolet emission, their study is particularly relevant to estimate the amount of high-energy irradiation onto the atmospheres of exoplanets, especially those in their stars' habitable zone.
*Aims.* We used the high-cadence, high-photometric capabilities of the CHEOPS and TESS space telescopes to study the detailed morphology of white-light flares occurring in a sample of 130 late-K and M stars, and compared our findings with results obtained at a lower cadence.
*Methods.* We employed dedicated software for the reduction of 3 s cadence CHEOPS data, and adopted the 20 s cadence TESS data reduced by their official processing pipeline. We developed an algorithm to separate multi-peak flare profiles into their components, in order to contrast them to those of single-peak, classical flares. We also exploited this tool to estimate amplitudes and periodicities in a small sample of quasi-periodic pulsation (QPP) candidates.⋆⋆
*Results.* Complex flares represent a significant percentage ($\gtrsim 30\%$) of the detected outburst events. Our findings suggest that high-impulse flares are more frequent than suspected from lower-cadence data, so that the most impactful flux levels that hit close-in exoplanets might be more time-limited than expected. We found significant differences in the duration distributions of single and complex flare components, but not in their peak luminosity. A statistical analysis of the flare parameter distributions provides marginal support for their description with a log-normal instead of a power-law function, leaving the door open to several flare formation scenarios. We tentatively confirmed previous results about QPPs in high-cadence photometry, report the possible detection of a pre-flare dip, and did not find hints of photometric variability due to an undetected flare background.
*Conclusions.* The high-cadence study of stellar hosts might be crucial to evaluate the impact of their flares on close-in exoplanets, as their impulsive phase emission might otherwise be incorrectly estimated. Future telescopes such as PLATO and Ariel, thanks to their high-cadence capability, will help in this respect. As the details of flare profiles and of the shape of their parameter distributions are made more accessible by continuing to increase the instrument precision and time resolution, the models used to interpret them and their role in star-planet interactions might need to be updated constantly.

## 1. Introduction

As an outcome of astrophysical dynamos, stellar activity manifests itself under a variety of timescales and wavelengths, and related observables are produced from the photosphere up until the corona. From the day- to month-, or even year-long cycles due to starspots, faculae, and their evolution (e.g. Hall 1991; Lanza et al. 1998; Namekata et al. 2019, 2020), down to impulsive events such as stellar flares and coronal mass ejections (CMEs) that are generated in active regions (e.g. Hudson 1991; Maehara et al. 2012; Walkowicz et al. 2011; Kowalski et al. 2013), these phenomena are among the main factors that shape the stellar energy balance and the environment where planets live. In this respect, the impact of flares and CMEs on planetary atmospheres, including their possible erosion, ionisation, and triggering of photochemical reactions in their upper layers, has recently attracted attention from many teams (e.g. Sanz-Forcada et al. 2010; Venot et al. 2016; Spake et al. 2018; Rodríguez-Barrera et al. 2018; Guilluy et al. 2020; Chen et al. 2021; Locci et al. 2022; Colombo et al. 2022; Maggio et al. 2022, 2023; Jackman et al. 2023; Louca et al. 2023). The interest in these topics was increased by the discovery of planets in the habitable zone of M stars (e.g. Gillon et al. 2017), whose high-energy irradiation is both able to stimulate the formation of biologically valuable molecules and to destroy the chemical bonds that make life as we know it possible (e.g. Rimmer et al. 2018; Barth et al. 2021). Another key parameter is the flare rate: frequent events are likely to increase their effect on exoplanet atmospheres, in addition to (when rocky) surface and even interiors, as they do not leave the time necessary for the induced perturbations to fade out (Venot et al. 2016; Vida et al. 2017; Hazra et al. 2020; Airapetian et al. 2020; Grayver et al. 2022; Nicholls et al. 2023).

---

⋆ This study used CHEOPS data observed as part of the Guaranteed Time Observation programmes CH_PR100018 (PI I. Pagano) and CH_PR100010 (PI G. Szabó). The CHEOPS photometry discussed in this paper is available in electronic form at the CDS via anonymous ftp to cdsarc.u-strasbg.fr (xxx) or via yyy.
⋆⋆ The main parts of the code we developed can be found on GitHub.

Flares on M stars are known to often reach several orders of magnitude higher energies than solar flares, which hardly go beyond $10^{32}$ erg (Maehara et al. 2012; Shibayama et al. 2013; Loyd et al. 2018). While flare emission extends from gamma rays to the radio, the most energetic events are those that can be observed in the optical, and happen when the concomitant extreme ultraviolet (EUV) and soft X-ray luminosity reaches particularly high levels (McIntosh & Donnelly 1972; Neidig 1983). Hence, the study of white-light flares, which behave similarly in the Sun and other stars (Neidig 1989), is key to understand the impact of stellar activity on planetary atmospheres and magnetospheres, including that of the Earth (e.g. Airapetian et al. 2016).

At the lowest energies, flares are not less interesting. Through reconstructions of the solar disc at about 1 min cadence in EUV wavelengths, it was possible to detect solar micro- and nanoflares (Shimizu & Tsuneta 1997; Aschwanden et al. 2000), defined by an emitted energy in the range $10^{24} - 10^{27}$ erg and $\lesssim 10^{24}$ erg, respectively. On other stars, such low-energy flares still lack observational confirmation (e.g. Yang et al. 2017). Assuming a similar flare generation mechanism, we would expect flares on M dwarfs to reach equally low energies. Despite the intriguing idea that these processes, if frequent enough, might provide a solution to the coronal heating problem (Parker 1988; Haisch et al. 1991; Hudson 1991), no evidence to date definitively supports nanoflares against competing hypotheses (Parnell & De Moortel 2012; Dillon et al. 2020; Bogachev & Erkhova 2023). In this respect, disc-integrated optical photometry does not provide conclusive information, as nanoflares are supposed to occur in the quiet corona and to be hidden in the optical measurement noise (e.g. Benz 2017, and references therein). This is one of the reasons why it is still unclear whether the distributions of the nanoflare and larger energy outburst follow similar trends (e.g. Maehara et al. 2015; Aschwanden 2022).

In single-filter, visible photometry, stellar flares have been observed down to cadences of minutes and tens of seconds. To date, the largest surveys of high-cadence white-light flaring stars have been carried out thanks to Transiting Exoplanet Survey Satellite (TESS) and the Next Generation Transit Survey (NGTS), with cadences of 20 s for the former (Howard & MacGregor 2022), and 10 s for the latter (Jackman et al. 2020, 2023). While increasing the time resolution from 30 to 1 min has not highlighted significant differences in the flare-emitted energy in Kepler data (Raetz et al. 2020), the energy output of the widely varied 20 second cadence flare profiles has only started to be explored.

The observation of M dwarf white-light flares at 20 s cadence revealed that a large fraction of outbursts has a non-classical profile (Howard & MacGregor 2022). Multi-peak flare shapes were found to be the norm, and could indicate cascades of emissions from a single active region or sympathetic flares from adjacent regions (Hawley et al. 2014; Davenport 2016; Schrijver & Higgins 2015; Kowalski et al. 2019). Such new data sets also allowed the detection of magneto-hydrodynamic (MHD) quasi-periodic pulsations (QPPs) in the stellar plasma down to a few minutes period, thereby offering new indications about their potential coronal heating role (Nakariakov & Melnikov 2009; Zimovets et al. 2021) and on their impact onto exoplanet atmospheres (e.g. Ramsay et al. 2021).

An encompassing picture of the physical avenues that trigger stellar flares requires detailed observations of the events occurring before the flare rise phase. Several authors have reported both photometric and spectroscopic reductions in the stellar flux before the impulsive phases of a few dMe stellar flares, with most of the increased absorption in the ultraviolet (UV, Rodono et al. 1979; Giampapa et al. 1982; Doyle et al. 1988; Peres et al. 1993; Ventura et al. 1995; Zalinian et al. 2002; Leitzinger et al. 2014). Such 'dips' or 'black-light flares' were found to happen within the half-hour prior to the flare impulsive phase, have amplitude from 1 up to 20% of the quiescent flux level, and duration from a few seconds up to a few tens of minutes. Attempted explanations connect these dips to details of the flare generation mechanism, and suggest these events might happen on the Sun, too (Henoux et al. 1990; Aboudarham & Henoux 1987; van Driel-Gesztelyi et al. 1994; Tovmassian et al. 2003). Stars with reported dips do not belong to the same sub-stellar type, as they span the M1 to M5 types, and are both isolated and in dMe binary systems. Additionally, dips do not anticipate particularly energetic flares: the one observed by Ventura et al. (1995), for example, emitted an energy of $\simeq 10^{32}$ erg in a no-filter optical bandpass, while superflares are characterised by energies of at least $10^{33}$ erg. However, the scarcity of observed events and the variety of characteristics of such dips hampered the possibility of comparing models and building a unified picture.

The Characterising ExOPlanet Satellite (CHEOPS) space telescope (Benz et al. 2021) has demonstrated exquisite photometric precision at 3 s cadence (Morgado et al. 2022), and so represents a unique opportunity to open an even newer window on stellar flares. In this study, we addressed the details of complex white-light flare morphology, looked for low-energy outbursts, searched for pre-flare dips and QPPs in a sample of 130 late-K and M dwarf stars observed as part of CHEOPS's ancillary science programme. We took advantage of the highest time cadence that the CHEOPS and TESS (Ricker et al. 2015) space telescopes can achieve, which enables a comparable precision in the detection of a few second to a few tens of second details in optical light curves, respectively. In Section 2, we describe our sample. In section 3, we discuss the details of our data sets and the reduction techniques we adopted. Section 4 is dedicated to the modelling of the flare candidate profiles and Section 5 to the experiments made to test the algorithm we developed for our analysis. In Sections 6 we present our results, which we discuss as we draw our conclusions in Section 7.

## 2. Target selection

Among the CHEOPS Guaranteed Time Observers operations, an Ancillary Science programme has been dedicated to monitoring the short-timescale micro-variability of main sequence late-K to M dwarfs. Some of our targets are stars which were considered for radial velocity exoplanet searches, for example, with the Calar Alto high-Resolution search for M dwarfs with Exoearths with Near-infrared and optical Échelle Spectrographs (CARMENES, Alonso-Floriano et al. 2015; Cortés-Contreras et al. 2017). Others were selected from the M dwarf-related literature: for instance, stars for which pre-flare dips were detected. The list of our 130 K5V to M5V targets, and their relevant parameters which are available in the literature, can be found in Table A.1.

We adopted the Gaia $G$ band (Gaia Collaboration et al. 2023) magnitude of the targets, and assessed their distance using the Early Data Release 3 measurements by Bailer-Jones et al. (2021). When available, stellar effective temperatures ($T_{\rm eff}$) were obtained from the PASTEL catalogue (Soubiran et al. 2010), otherwise from a more general literature search and comparison with the spectral type reported on SIMBAD (Wenger et al. 2000).

We also searched the literature for activity indicators. We inspected the Strasbourg astronomical Data Center (CDS)[1] for published values of X and UV emission, $\log R'_{\rm HK}$, and $v \sin i$, and found this latter parameter to be the most frequently available (122 out of 130 targets). When different $v \sin i$ values were reported, we adopted their mean, and found $v \sin i < 5$ km s$^{-1}$ for 80% of the objects. Among those targets with an available $\log R'_{\rm HK}$ (33 out of 130), we found 67% to have a value $< -4.8$. As we were only interested in a general indication of the activity level of our targets, we did not correct the $\log R'_{\rm HK}$ values for interstellar extinction. We did not attempt to measure stellar rotation periods from TESS data, as the 28-days duration of TESS light curves is likely too short to capture a full rotation for most of the late-type stars in our sample (McQuillan et al. 2014).

Histograms for the retrieved stellar parameters are illustrated in Figure 1, and are consistent with a mostly low activity level for our objects. This was expected, given that most of our targets were selected from exoplanet-search stellar samples.

## 3. Observations and data reduction

The main steps of our flare search were 1) detrending of each light curve to evaluate the 'quiet' stellar flux level; 2) identification of flare peak candidates; 3) flare validation and profile fit; 4) comparison of the flare profile fit with a model including a pre-flare flux drop and inspection for QPPs in the fit residuals. The analysis was carried out with an automatic algorithm that we developed for this specific purpose. We performed a visual inspection of the results, which helped us tune the parameters of our algorithm to better suit the different types of data. In both cases, we found steps 1) and 3) to be the most critical and the most prone to errors.

### 3.1. CHEOPS light curves

This paper reports on the full extent of the ancillary science programme dedicated to late-K and M dwarf flares (CH_PR100018, PI I. Pagano), active between the beginning of scientific observations in April 2020 and September 2023, totalling ∼ 95 days spent on target. Further AU Mic observations from programme

[1] http://cds.unistra.fr/

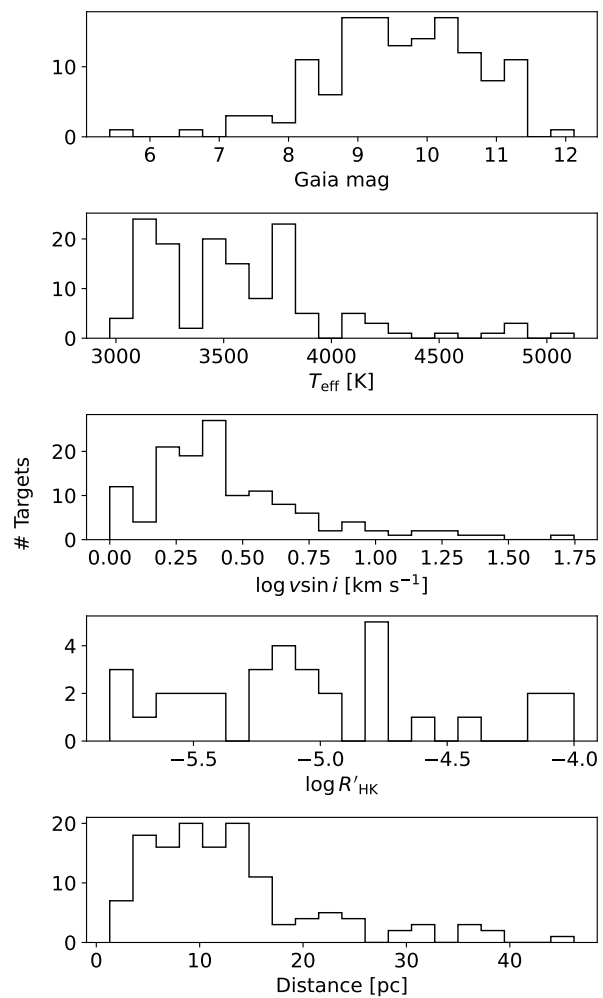

Fig. 1: Parameters for our stellar sample. From top to bottom: histograms for the Gaia magnitude, published $T_{\rm eff}$, $v \sin i$, $\log R'_{\rm HK}$ values and distance.

CH_100010 (PI G. Szabó) were used, for an additional ∼ 11 days observation time.

To study the detailed morphology of flares, photometric precision is as important as time resolution. Our goal was to obtain a flux measurement every few seconds; however, the CHEOPS standard mode of operations for $V \gtrsim 9$ stars (i.e. most of our targets) delivers images called 'subarrays' every few tens of seconds, as a result of on-board stacking of shorter exposures. These shorter exposures are not lost, but are recorded and downlinked as 50-pixel diameter 'imagettes' (Benz et al. 2021). Working with imagettes involves a partial loss of information compared to subarrays, which have a 200-pixel diameter (e.g. about each exposure's background), so that these latter are still useful for the correction of instrumental effects. The imagette exposure time can be adjusted and affects the subarray exposure time: for our operations, we found a 3 s cadence to be a good compromise between the actual on-target time and the telescope duty cycle. CHEOPS can reach as low a cadence as 1 ms, but this is only useful for $V < 6$ targets and would have implied a suboptimal combination of on-target time and duty cycle for our targets.

Imagette data sets are not automatically processed by the Data Reduction Pipeline (DRP, Hoyer et al. 2020), so that we relied on an ad hoc tool for imagette reduction: the Data Reduc-tion Tool (DRT), described in Morgado et al. (2022) and Fortier et al. (2024). After processing and for every visit, we rejected those points with a background flux deviating by more than $5\sigma$ from its median value, and clipped the measurements with a flux value $> 5\sigma$ lower than a smoothed version of the light curve. In order to avoid removing data points belonging to flares, we did not carry out any clipping above the smoothed flux level.

We normalised the light curve of each visit with a low-order polynomial as a function of time, so to remove any long-period trend likely due to stellar activity, with GJ 65 and AU Mic being the only targets which required a higher degree. We tested polynomials with degrees from three to ten and, for each degree, iteratively sigma-clipped the fit residuals until no more residual data points were above the clipping threshold. This allowed us to minimise the impact of the flares in the detrending procedure. After this, the polynomial fit with the lowest Akaike Information Criterion (AIC, Akaike 1974) was adopted.

Because of the irregular shape of CHEOPS's point spread function and the nadir-locked rotation of the telescope aimed at maintaining thermal stability, the satellite's observations are affected by contamination from nearby stars which are in phase with its $\varphi \in [0, 2\pi)$ roll angle (e.g. Lendl et al. 2020). To detrend the data from this instrumental effect, we combined all observations for a given target to maximise their signal-to-noise ratio (S/N) in the roll-angle space. Given the constraints imposed by the filler nature of the programme, we found a duration of three CHEOPS orbits (i.e. $\sim$ 300 min) for each observing sequence or 'visit' to be an affordable compromise. After the first months of operations, we therefore set this duration for all the visits of our programme. Also, merging all visits for a given target minimised the data gaps due to the passage of the telescope across the South Atlantic Anomaly (Benz et al. 2021), and amounting to up to 50% of each visit observation time.

We then followed Scandariato et al. (2017) to fit a model $\Theta$ for roll-angle systematics with a combination of sines and cosines, that is,

$$\Theta(\varphi) = \sum_{i=1}^{5} \left[ a_i \sin(i \cdot \varphi) + b_i \cos(i \cdot \varphi) \right], \quad (1)$$

where $a_i$ and $b_i$ are the $i$-th coefficients in the Fourier reconstruction of the sinusoidal signal. After experimenting with the number of harmonics, we found $i$ up to 5 was able to provide sufficient flattening of the residuals. The standard deviation of the flattened light curve, without outliers and flare candidates, was used as the data point uncertainty in the flattened light curve. The result of the detrending for the light curves of GJ 65, where both instrumental effects and stellar long-period variability were removed, is shown in Figure 2.

Our detrending procedure was effective for most datasets, but less so for the few light curves that were contaminated by a Solar System body lying within $\lesssim 25°$ from the target position, or by particularly bright neighbouring objects. The resulting strong non-periodicity of the roll-angle signal in different visits required a visual inspection of the detrended light curves, the masking of some portions of the data, and the discarding of false positive detections. In particular, we rejected the light curves of 27 out of 130 targets (21%), where however we could not identify any evident flare by visual inspection.

### 3.2. TESS light curves

The TESS telescope uses a similar, even if redder, photometric band to CHEOPS[2] Its full-sky coverage, photometric precision, and 20 s cadence mode made available since its Cycle 3 (Sector 27 onward), was relevant for our goals, as it allowed long, continuous, and high-cadence observations of several of our targets. We therefore downloaded from the Mikulski Archive for Space Telescopes [3] all available 20 s TESS light curves for our targets at the time of writing, or 106 light curves for 73 objects, until Sector 67. The list of programmes, and the corresponding PIs, from which the data were obtained can be found in Table A.2.

We used the Pre-search Data Conditioning Simple Aperture Photometry (PDCSAP) flux, where instrumental trends are removed by the TESS pipeline (Jenkins et al. 2016). As reported in a few cases, PDCSAP correction might add spurious effects to the data (e.g. Nardiello et al. 2022); in our case, however, the trend and flare timescales are different enough to prefer this latter over the Simple Aperture Photometry (SAP) format, where instrument artefacts are not removed.

We first derived a flattened version of each light curve by removing its respective smoothed version. The smoothing function was computed with a Hann window, with a window length of $\sim$ 0.2 times the duration of the most likely periodicity in the light curve. This latter was found using ASTROPY's implementation of the Lomb-Scargle periodogram (Astropy Collaboration et al. 2013, 2018, 2022). The smoothing was carried out iteratively, by sigma-clipping the outliers from the smoothing process at each iteration and stopping when no more outliers were present, and therefore minimising the contribution of the flares to the detrending process.

We also tested other approaches to smooth the light curves, such as pre-whitening and Gaussian Process regression. In all cases, we found that the smoothing resulted in incorrectly increased smoothing function values before and after the most energetic flares, so that their flattened profiles were dampened. This also caused spurious pre- and post-flare flux drops to appear. We took this into account by determining correction factors on our results based on injection tests (Section 5) and applying a double-validation test to any fitted pre-flare dip candidate (Section 4.2).

Figure 3 shows the data median absolute deviation (MAD) on the CHEOPS subarray and detrended CHEOPS imagette light curves, as well as on the flattened 20 s TESS light curves (the CHEOPS subarray data were reduced by its dedicated DRP). The CHEOPS imagette light curves were resampled in 20 s intervals for the sake of comparison: the Figure shows that the average MAD of the detrended CHEOPS imagette light curves, even if higher than the MAD of the CHEOPS DRP data, is still lower than the one achieved by TESS on our targets.

## 4. Flare detection and validation

Flare candidate peaks were searched in the flattened light curves with the PEAKUTILS software (Negri & Vestri 2017), which relies on the inspection of the first derivative of an array, and where we set a minimum peak threshold of four times the noise value. This threshold was chosen following several experiments and visual inspection of the results: it was indeed found that a threshold of $5\sigma$ would yield a significant number of missed low-amplitude flares, while a $3\sigma$ threshold would produce a large number of false positives, such as clear statistical fluctuations or outliers that were flagged as flares. While most false positives could have been rejected during the flare validation phase, we preferred avoiding the risk of increasing their number. Following visual inspection, we also set a minimum duration of five and three data points for CHEOPS and TESS candidates, respectively, in order to limit the number of false positive detections. If a) the so-defined flare lasted more than the assigned minimum duration, b) no part of the flare profile fell in a data gap, and c) the first two points after the peak candidate were at least $2\sigma$ above the noise level, we proceeded with the fit of the flare profile. To do this, we computed a moving median-filtered stellar flux with a window dependent on the data cadence (11 points for TESS data, 29 for CHEOPS); then, we considered the stellar flux from one hour before the time when the median-filtered flux exceeded the noise level, to one hour after the time when the same flux returned to the noise level.

---

[2] A comparison of their bandpasses can be found at this page.
[3] https://archive.stsci.edu/tess/bulk_downloads/bulk_downloads_ffi-tp-lc-dv.html

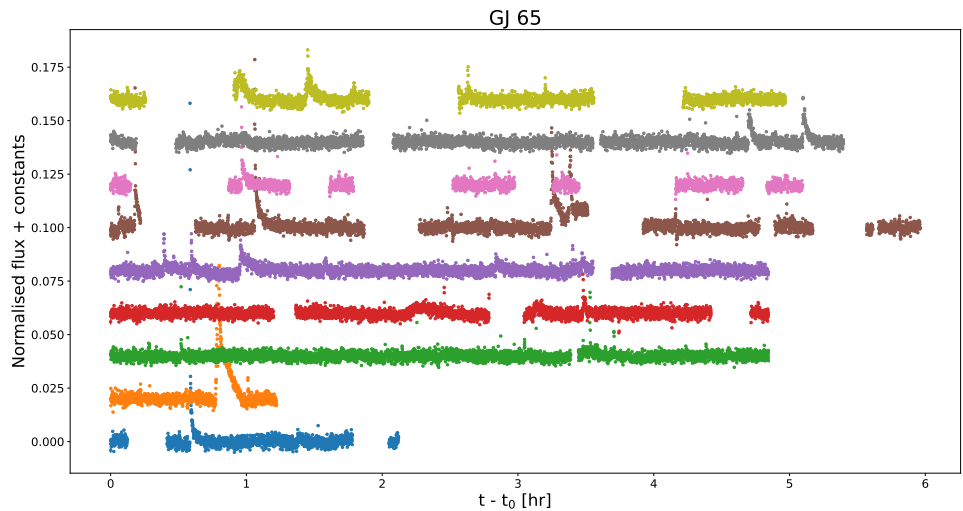

Fig. 2: GJ 65 CHEOPS DRT imagette light curves, detrended for roll-angle effects and stellar variability. Each CHEOPS visit is assigned a different colour and shifted for clarity.

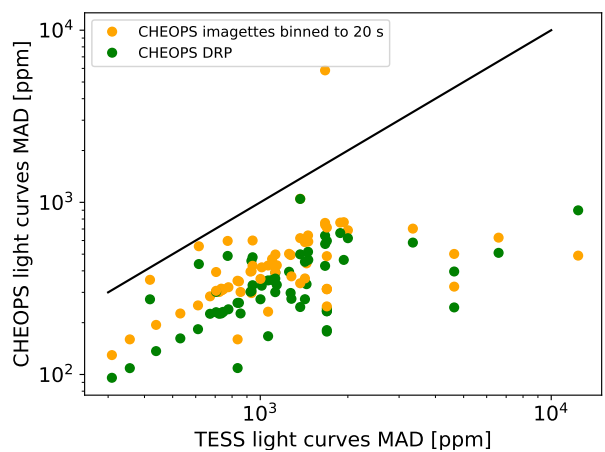

Fig. 3: Median absolute deviation (MAD) of the CHEOPS subarray (DRP) and imagette (DRT) light curves (binned to 20 s), as a function of the one of the 20 s TESS light curves. Only targets with both CHEOPS and TESS data are represented. The one-to-one line is marked in black.

### 4.1. Flare model and fit

Flares sampled at high cadence often deviate from templates found at lower time resolution, and present multiple peaks, flat-top peaks, Gaussian-shaped bumps and oscillations (e.g. Howard & MacGregor 2022). While analytical models can hardly capture the full complexity of these profiles, they are nonetheless useful to test the accuracy of empirically derived templates, and can be used to validate them and to better identify the quiet stellar flux level.

We modelled each flare profile $f(t)$, $t$ being time, following the double exponential decay that Davenport et al. (2014) found to be representative of most white-light flares observed by Kepler on GJ 1243. This model separates a first, linear decay phase (possibly explained by bremsstrahlung radiation), and a second, gradual phase (probably connected to radiative recombination). Additionally, we smoothed the flare peak following Gryciuk et al. (2017), Jackman (2020), and Mendoza et al. (2022)'s prescription, by taking the convolution of the double exponential with a Gaussian function to represent both an energy release and loss process. This model, which simulates a gradual transition from the rise to the decay phase, proved to be particularly useful to describe the classical exponential shape, the 'bump' and partially the 'flat-top' flare components (e.g. Howard & MacGregor 2022) with the single functional form

$$f(t) = \frac{\sqrt{\pi}AC}{2}\left[(F_1 h(t, B, C, D_1) + F_2 h(t, B, C, D_2)\right], \quad (2)$$

where

$$h(t, B, C, D) = \exp\left[-Dt + \left(\frac{B}{C} + \frac{DC}{2}\right)^2\right]\mathrm{erfc}\left(\frac{B-t}{C} + \frac{DC}{2}\right), \quad (3)$$

and where $\mathrm{erfc}(t) = 1 - \mathrm{erf}(t)$ is the complementary error function. In the above expressions, $A$ is the normalised peak amplitude of the flare, $B$ is the peak time in a rescaled time axis (see below), $C$ its rise timescale, $D_1$ and $D_2$ the fast and slow cooling timescales, and $F_1$ and $F_2$ (with $F_1 = 1 - F_2$)) the relative importance of the two cooling phases.

For each flare candidate, we carried out a least-square fit with the Powell minimisation algorithm implemented in scipy (Virtanen et al. 2020), and wrapped in the LMFIT package (Newville et al. 2014). A more thorough exploration of the parameter space was made possible thanks to the basin-hopping optimisation algorithm (Wales & Doye 1997), available in the same library. The method also provides estimates for the parameter uncertainties using the fit covariance matrix.

To improve convergence, we fixed the model parameters to the values Mendoza et al. (2022) fitted on a set of single-peak GJ 1243 flares, and fitted only the flare peak time $t_{peak}$, its amplitude (which scales the $A$ term) and its full width at half maximum (FWHM). The time axis $t$ in Equation 2 was then rescaled as $t' = (t - t_{peak})/\text{FWHM}$.[4] To model any residual trend in the quiet stellar flux level, a first-order polynomial was included in each flare profile fit.

A final flare validation was performed based on the AIC value of a one-peak flare model against a second order polynomial fit to the flux centred around the candidate peak: if the AIC favoured this flare fit by at least six units, that is, if the model without flares was < 5% as probable as the model with one flare to minimise the information loss from the data, the flare was considered validated.

Every validated flare was examined for the presence of additional peaks by adding one flare profile at a time, up to a maximum of five flares. If the AIC of a fit with more flares was preferred to the model with less flares, in a similar way to Davenport et al. (2014), the new fit was retained. Once again, the requirement for the more complex model was set to at least six units a reduction in the AIC value.

To help prevent outliers in complex flare profiles from being misinterpreted as flares, we fixed a lower bound to the fitted flare FWHM for the flare to be validated. We found validating flares with this criterion to be more effective than setting a lower bound to this parameter during the fit, as in the second case the fits of very short-duration flares tended to hit the lower parameter bound. We then required for each candidate flare to have a fitted FWHM larger than the data cadence, that is, 3 s and 20 s for the CHEOPS and TESS sample, respectively. Also, to avoid a similar effect in amplitude, we rejected flares with amplitude smaller than 1.1 times the lower bound value.

*4.2. Pre-flare flux dips*

We used the same method to inspect for the presence of pre-flare flux drops that might be indicative of black-light flares. For each validated outburst, we searched the flux within one hour prior to the rise phase, and fitted a generalised Gaussian model to it:

$$f(t) = \begin{cases} G\exp[-(|t - t_d|/w_1)^n] + q & \text{for } t < t_d, \\ G\exp[-(|t - t_d|/w_2)^n] + q & \text{for } t \geq t_d, \end{cases} \quad (4)$$

where $G < 0$ is the Gaussian lowest flux value, $t_d$ is the Gaussian function central time, $w_1$ and $w_2$ represent the Gaussian width on each side, $n > 2$ allows the function to assume less peaked shapes than a standard Gaussian, and $q$ allows for the quiet stellar flux level to be adjusted. We attempted to reduce the dependence on the initial condition for $t_d$ by using a basin-hopping optimisation with 10 iterations.

We then compared the AIC of the best fit with the one of a flare without flux drops. As a first validation test against the flare-only model, we retained the dip fit if its AIC was lower than the one without it by at least six units. We also noticed a tendency for the fit to find dip features in data sets with a correlated structure in the quiet stellar flux level, which might be a residual from the detrending or from systematic effects. We exclude stellar granulation as a possible explanation, because its signal is too faint to be noticeable in a single M star TESS (Sulis et al. 2023) or CHEOPS light curve. We then compared the amplitude of the fitted flux drop with the correlated noise level in the light curve after detrending or smoothing. For each CHEOPS target or TESS light curve, this was calculated with Pont et al. (2006)'s method, and the flux drop was compared to the noise level corresponding to the drop's fitted width. We considered a $2.5\sigma$ flux drop significance against the red noise level as a threshold for a dip candidate to be visually inspected.

---
[4] An example of such an implementation can be found at this GitHub page.

## 5. Injection tests

To test the accuracy of our detection and fit algorithms, we ran injection tests on the CHEOPS and TESS light curves without detected flares. To create a statistically meaningful sample, we used 100 TESS and 500 CHEOPS light curves, so that some CHEOPS data sets were used for the test more than once. Using the model presented in the previous sections, we added between one and two flares to each CHEOPS visit, and between 10 and 50 flares at random times to each TESS light curve, with log-uniformly distributed amplitudes corresponding to S/N between 1 and 20, and a log-uniformly distributed FWHM between 3 s and 100 min. We also added pre-flare dips before each flare peak, with amplitude between 0.05 and 0.5 times the S/N of the associated flare. The scaling between the S/N of the flare and of the associated dip was adopted to reduce the likelihood of missing large dips following low-amplitude flares. Dips were allowed to have asymmetric widths between 0.1 and 5 min.

We ran our algorithm on the flare-injected light curves after a previous detrending phase, applying the same algorithms described in Sections 3.1 and 3.2. A simulated flare or dip was considered recovered if the algorithm validated a flare within 3 min before and after its true peak time. Figure 4 presents the recovery rates of flares and dips as a function of their amplitude: the recovery rate was defined as the ratio between the number of validated and injected events in a given amplitude range, and the false positive rate as $n_{false}/(n_{true} + n_{false})$, where $n_{false}$ and $n_{true}$ are the correct and incorrect detection numbers, respectively. In terms of dips, we only show those for which the AIC prefers the model including the flux drop by at least 6 units, and with a significance of at least 2.5 when compared to the correlated noise level of the light curve according to Pont et al. (2006)'s method.

Figure 4 shows that the completeness of our detection algorithm approaches unity for TESS flares with amplitude close to $\sim 10\%$, while it is lower for CHEOPS. This can be explained by a combination of poorly corrected systematics and the fact that the largest flares are partly removed by the CHEOPS detrending algorithm, once their duration becomes comparable to the duration of the CHEOPS visits ($\simeq 3$ hours).

Similarly, the completeness of dip detections is higher for TESS than for CHEOPS data. This can once again be explained by both the detrending procedure for the CHEOPS data, which often removes the largest dips, as well as the fact that dips might fall into data gaps.

In Figure 5, we present the retrieved flare amplitude, FWHM, and dip amplitude against their injected values, for the correct detections. For all pair of retrieved $p_{ret}$ and injected $p_{inj}$ parameters, we fitted quadratic polynomials with the form

$$\log p_{inj} = c_1 \log^2 p_{ret} + c_2 \log p_{ret} + c_3 \quad (5)$$

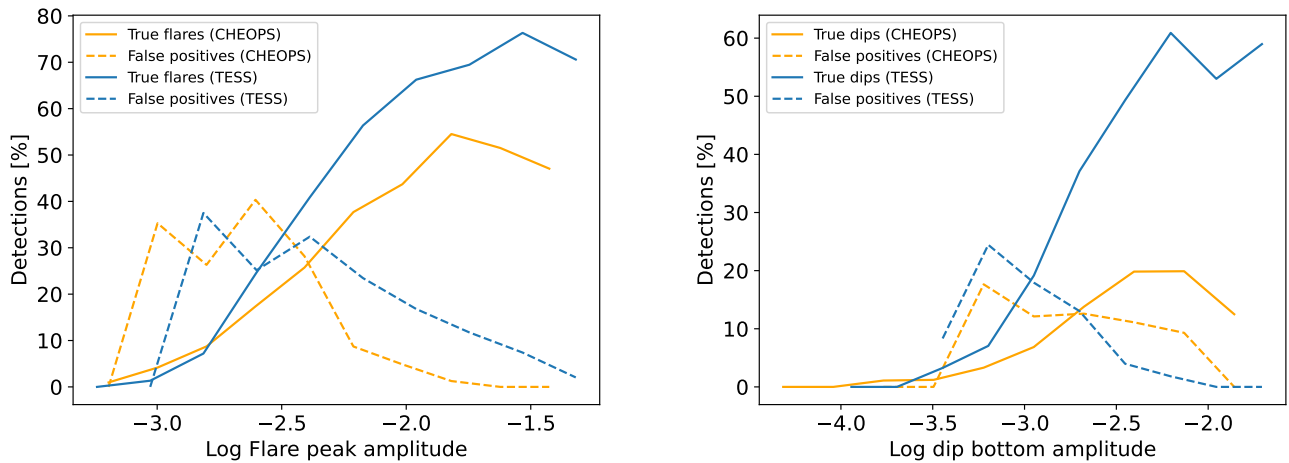

Fig. 4: Correct detections and false positives for flares (*left*) and pre-flare dips (*right*). The results for CHEOPS and TESS simulations are shown with different colours.

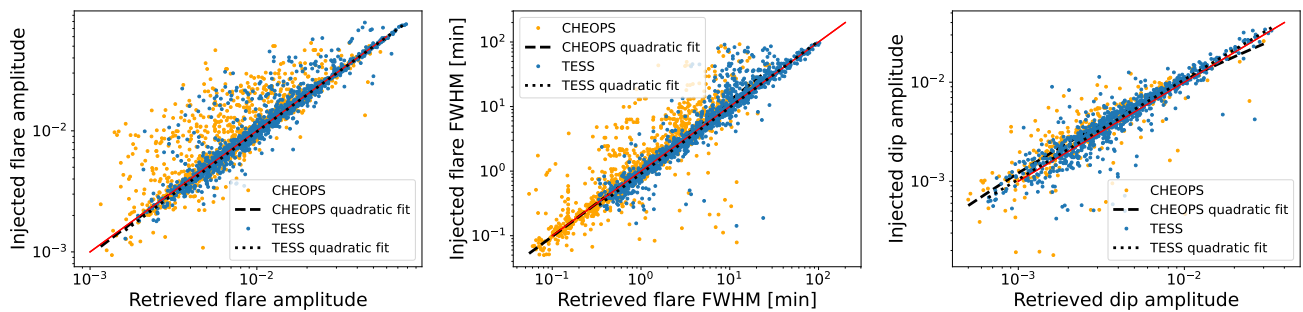

Fig. 5: Retrieved versus input parameters used in the injection tests. The one-to-one relationship is highlighted with red lines, and quadratic polynomial fits for the CHEOPS and TESS data are plotted as dashed and dotted black lines, respectively.

|  | Flare amplitude | Flare FWHM | Dip amplitude |
|---|---|---|---|
| **CHEOPS** | | | |
| $c_1$ | −0.0302(1) | 0.01863(8) | −0.116(2) |
| $c_2$ | 0.877(3) | 1.05247(3) | 0.36(5) |
| $c_3$ | −0.120(3) | 0.02155(2) | −0.80(7) |
| **TESS** | | | |
| $c_1$ | −0.01808(7) | −0.00032(3) | −0.0069(4) |
| $c_2$ | 0.947(1) | 1.02639(8) | 0.973(9) |
| $c_3$ | −0.041(1) | −0.03846(1) | 0.00(1) |

Table 1: Coefficients for the fitted relationship between retrieved and injected flare parameters, for simulated CHEOPS and TESS data. Numbers in parentheses indicate the uncertainty on the last reported digit.

to the simulated CHEOPS and TESS data separately, and obtained correction factors that we later applied to the results on the observed data. Flare durations, defined as the time during which the fitted flare model is above the photometric noise level, were re-estimated by recomputing each flare profile with the corrected amplitude and FWHM. The $c_1$, $c_2$ and $c_3$ coefficients are presented in Table 1.

## 6. Results

### 6.1. Flare rate

We validated 100 and 1364 flares in the CHEOPS and TESS light curves, respectively, considering the components of multi-peak flares as individual events. The left panel of Figure 6 compares the number of non-flaring and of flaring stars as a function of stellar spectral type. Most detected flares were observed for the latest spectral types, as expected from previous results (e.g. Günther et al. 2020; Jackman et al. 2021).

Comparing targets with both CHEOPS and TESS observations, we found a flare rate in the range 0-15 day$^{-1}$ for objects observed with the former, and between 0 and 4 day$^{-1}$ for those observed with the latter. While these results are broadly compatible with the flare rates found by Günther et al. (2020) on a sample of 24809 M dwarfs observed with 2 min cadence during the first two months of the TESS mission, we refrain from a direct comparison because of the different size of the CHEOPS and TESS validated flare sample.

The validated events here presented are those that passed the fitted duration and amplitude criteria outlined in Section 4.1. However, our results might still be affected by false positive detections due to instrumental correlated noise, especially because some of the validated flares have a very low amplitude. Given the results of our injection tests, we inspected the likely contribution of false positives in our results. By reference to the left panel of Figure 4, we used as thresholds the flare log-amplitude values above which the detection rate becomes higher than the false positive rate: about −2.55 and −2.4 for TESS and CHEOPS, respectively. The flares with fitted amplitudes below these thresholds are 3% for TESS and 35% for CHEOPS. We therefore carried out a visual inspection of the CHEOPS flares validated by our pipeline, to remove obvious false positives: 26 validated flares were rejected. For this reason, the percentage of CHEOPS false positives here presented is much lower than the one obtained from the tests.

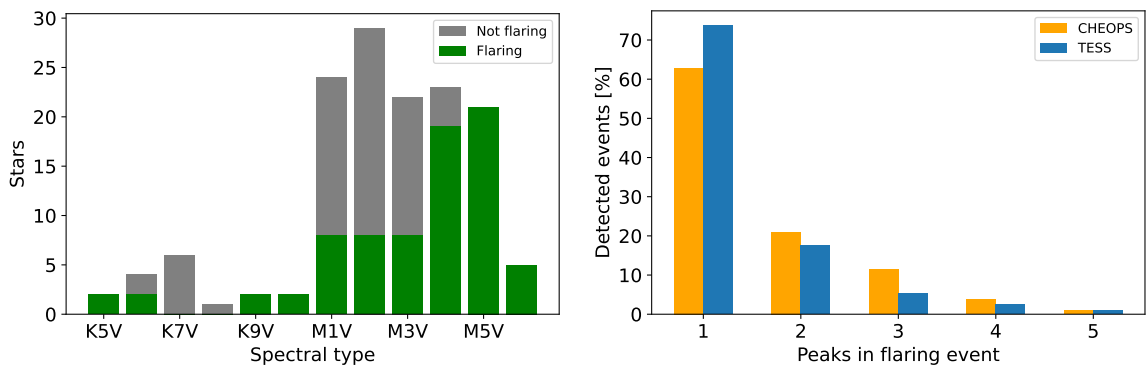

Fig. 6: Flare detection statistics. *Left:* Number of flaring and non-flaring stars as a function of spectral type. *Right:* Percentages of peaks observed per flare event. Flares observed with CHEOPS and TESS are distinguished.

### 6.2. Flare complexity

A combination of Mendoza et al. (2022)'s smoothed, two-phase templates, whose average shape was obtained on a set of single-peaked flares on GJ 1243, was found to successfully reproduce most complex flare components, too. The 3 s cadence data analysis suggests that the higher the cadence, the higher the complexity of flares that might be revealed, with an increase in the fraction of three-peak flares compared to 20 s observations. As shown in the right panel of Figure 6, about a third of the overall fitted flares presents more than one peak. This confirms the relative importance of complex flares once the observing cadence can resolve their structure (Howard & MacGregor 2022). In particular, many of the two-peak flares are of the peak-bump type discussed by these authors, even if we detected cases where the 'bump' precedes the 'peak'. We also found a few occurrences of structure in the rise phase and a flat-top profiles. Examples for such cases are shown in Figure 7.

We fitted at most five individual components in multi-peak outbursts, but found by visual inspection flares with an even more complex structure. While fits with a larger number of components could be attempted, care should be exercised against the increasing number of degeneracies and correlations among the parameters, as well as dependence on the minimisation algorithm. However, our results suggest that particularly complex flares are a minor component in our sample of mostly low-activity stars ($\lesssim$ 5%). This number accounts also for the QPP candidates that we observed (Section 6.7): to be conservative, such candidates were removed when deriving the flare properties discussed in this Section. In this regard, we remark that an estimate of a few percent of QPPs in the flare sample is in agreement with the M star literature (Ramsay et al. 2021; Million et al. 2021; Howard & MacGregor 2022).

According to Tovmassian et al. (2003), the peak-bump shape is the fundamental flare shape (even preceded by a starspot-induced pre-flare dip), as energy is re-radiated by the stellar photosphere after the peak phase. However, the complete profile is observable only when the emission site is close to the centre of the visible stellar disc. In this hypothesis, flares that occur near the stellar limb have their 'bump' part less visible, but if they are particularly powerful and their peak occurs in the hidden part of the stellar disc, we can only see a residual 'bump'. As the bump would be visible from the largest fraction of the stellar hemisphere, these authors claim flares with sharp rise and decline but no bump should be about three times less abundant compared to the peak-bump type.

We could not directly test this hypothesis, as our implementation does not clearly separate 'sharp' and 'smooth' single-peak flares. We could, nonetheless, attempt a distinction between these two flare types through the distribution of their impulses, given by the ratio between their peak amplitude and FWHM (e.g. Howard & MacGregor 2022). If Tovmassian et al. (2003)'s model is correct, then the first event in each two-peak flare should have relatively high impulse, as it would correspond to the fast rise-fast decay part of the peak-bump flare prototype. The impulse of single-peaked flares should instead be more evenly distributed between both low and high values, which roughly represent the 'bump-only' and the 'peak-only' type, respectively. Figure 8 shows the distributions of the impulse for these two types of flares: on average, main events in flare pairs have larger impulse values, and the null hypotheses of the two distributions belonging to the same one is rejected by a Kolmogorov-Smirnov (KS) test with $p$-value of 0.003. However, the distribution of single-peak flares is not evenly distributed among all values: this might be related to the little likelihood of finding a powerful enough flare which leaves its 'bump' imprint right behind the stellar limb.

To conclude, we inspected the waiting time in complex flares, that is, the time between peaks of consecutive individual components (e.g. Hawley et al. 2014). We found most events to happen a few minutes after a preceding outburst, as shown in Figure 9. This finding recalls solar sympathetic flares, namely pairs of flares which happen in distinct but connected active regions. There, the interval between twin flares can last up to a few hours, depending on the solar cycle phase (Mawad & Moussas 2022).

### 6.3. Simple and complex flare morphology

Hawley et al. (2014) divided single- and multi-peak flares in the study of their parameters. They found that simple and complex flares have a similar amplitude distribution, but that complex flares tend to be longer and more energetic. However, their result might be affected by the fact that they did not separate the individual components of complex flares. We instead distinguished actual single-peak flares from the components of complex outbursts, while taking care of the fact that the comparison of flare amplitudes obtained for different stars is biased by different stellar magnitudes. Following Raetz et al. (2020), we converted flare amplitudes to flux by adopting the CHEOPS and TESS zero-points and effective wavelengths provided by the filter profile service of the Spanish Virtual Observatory (Rodrigo et al. 2012). By using the stellar distances and magnitudes (Section 2), we obtained the luminosities. For CHEOPS data, we used the targets' Gaia $G$ magnitudes, and for TESS data the TESS magnitudes available in each data set fits file.

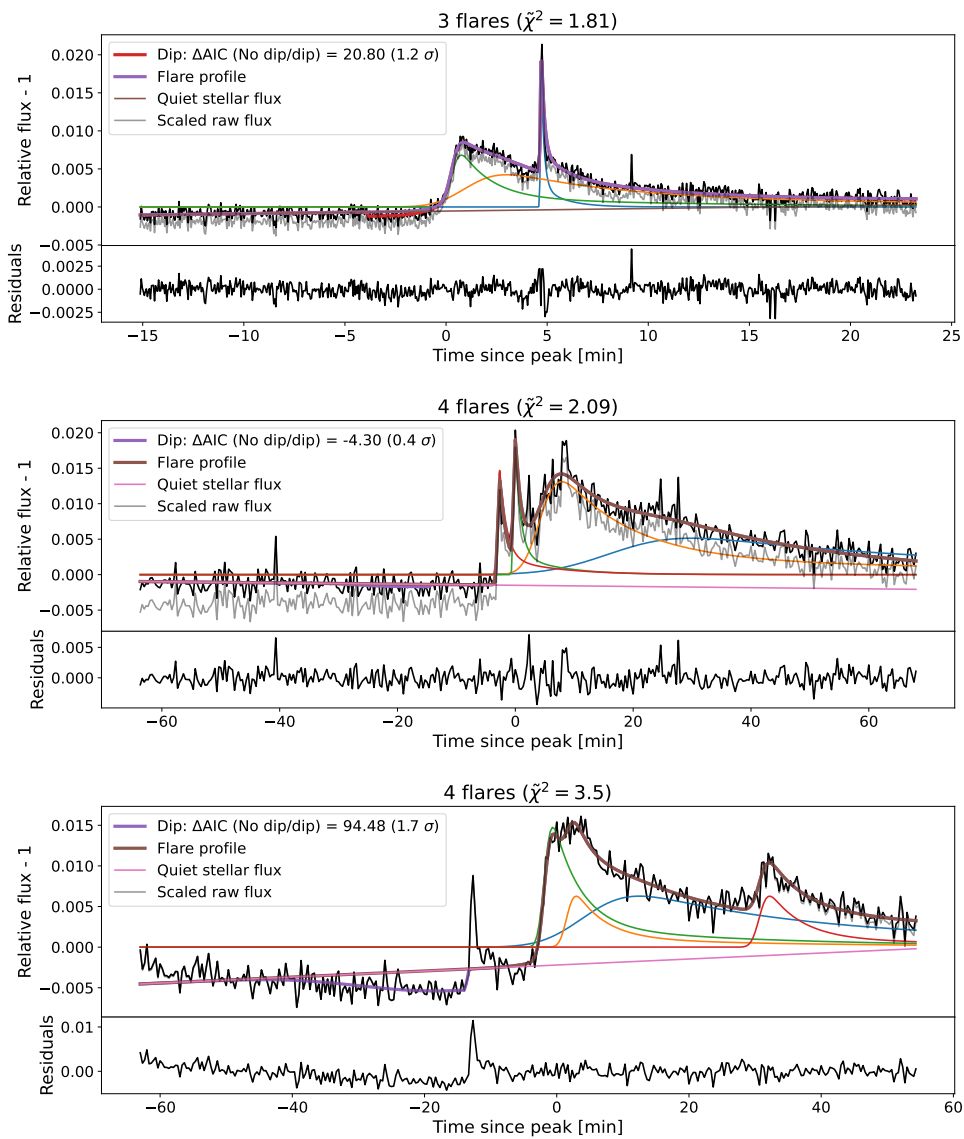

Fig. 7: Examples of complex flare profiles. *Top:* A bump-peak flare profiles on V1054 Oph. Individual flare components are represented with different colours, and the total model (including the quiet stellar flux) is drawn with a thicker line. The grey, half-transparent line shows the raw flux before light curve detrending. The legend reports the AIC difference between a model without and with dip, and the fitted dip significance with respect to the correlated noise level. The rapid flux drops after min. 15 are likely not to be attributed to the stellar signal. *Centre:* A flare with structure in the rise phase observed on YZ Cmi. *Bottom:* A flat-top profile flare on GJ 799B. Our fit strives to model it with a single profile and associates two flares to it. As we found such profiles to be rare, they do not significantly affect our complex flare statistics.

Once flare luminosity is considered and complex outbursts are separated in their individual components, not all parameters of single and multi-peaked events are indeed distributed in the same way, as illustrated in Figure 10: a KS test on the two luminosity distributions cannot distinguish them with a $p$-value of 0.86, while the duration distributions are distinguished with $p \sim 10^{-49}$.

It was also previously reported, using *Kepler/K2* data (e.g. Hawley et al. 2014; Raetz et al. 2020), that white-light flare luminosity $L$ and duration $d$ are linearly correlated for flares with duration between $\sim 5$ and $\sim 100$ min. The coefficients of this linear relationship depend on the time cadence and the noise level of the data, as found on Kepler flares observed simultaneously in long and short cadence (Yang et al. 2018). In particular, a higher cadence allows the resolution of higher peak values, and a lower noise level enables the detection of longer flare durations. The two effects were found to compensate in K2 when integrated energy levels are computed (Raetz et al. 2020).

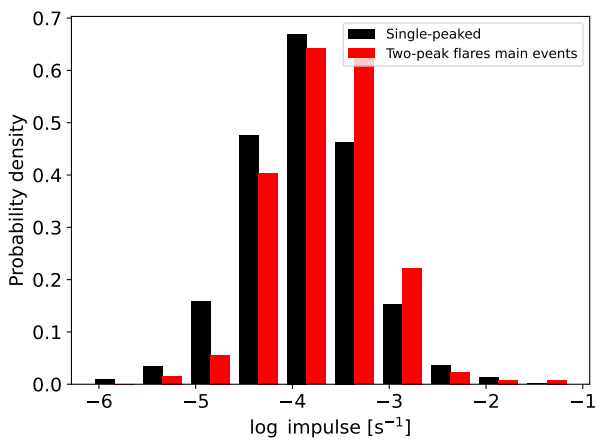

Fig. 8: Impulse for single-peaked flares (*black*) and the first component of two-peaked flares (*red*).

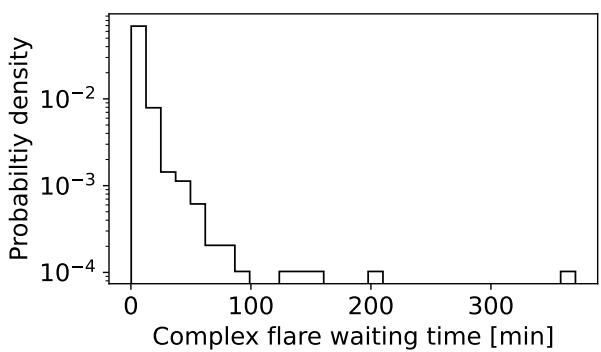

Fig. 9: Distribution of waiting time between consecutive flare components.

Even if the flare *L-d* trend is instrument-dependent, it can be informative if comparisons are made among subsets of events observed in the same setting. We confirm the trend holds when including higher-cadence data: in particular, we found a Pearson correlation coefficient of 0.46 ($p$-value for non-correlation $\sim 10^{-56}$) and 0.28 ($p \sim 10^{-8}$) for simple and complex flare components, respectively. We estimated the coefficients of an *L-d* linear relationship by bootstrapping 1000 times every validated flare's amplitude and FWHM based on their mean value and uncertainty; $L$ and $d$ were re-calculated at each iteration from the corresponding profile. The fits revealed two different slopes for the simple and complex flare trends, which might be explained by their different duration distributions:

$$\log L_{\text{simple}} = (0.36 \pm 0.01) \log d_{\text{simple}} + (29.17 \pm 0.02) \quad (6)$$

and

$$\log L_{\text{complex}} = (0.48 \pm 0.01) \log d_{\text{complex}} + (29.27 \pm 0.01), \quad (7)$$

where luminosity and duration are measured in erg s$^{-1}$ and minutes, respectively. The measurements and fits are shown in the left panel of Figure 11. Some of the validated flares had very small uncertainty on one or both amplitude and FWHM: as this was probably a numerical effect due to the minimisation algorithm, we reduced the weight of these events by assigning a relative uncertainty of 10% to all parameters with relative uncertainty $< 10^{-2}$ on any of these parameters (5% of the flares).

The fitted coefficients for the *L-d* relationship are significantly lower than the ones found by Raetz et al. (2020), who reported a linear coefficient of $0.93 \pm 0.04$ on the K2 short cadence ($\simeq 1$ min) data of M stars with a wide variety of activity levels: this can be explained by the differences among Kepler, CHEOPS and TESS. Interestingly, Brasseur et al. (2019) did not find a significant correlation between flare luminosity and duration, using a sample of 1904 flares observed at 10 s cadence with GALEX near-UV.

Constraining the *L-d* relationship might help refine the models used to compute the energy flares can deposit onto exoplanet atmospheres, and particularly to scale the prototypical flare profiles that are used (e.g. Nicholls et al. 2023). In this regard, the flare FWHM is a proxy for the time during which the emission rises and decays more rapidly. For example, Venot et al. (2016) modelled the flare-induced chemical perturbations that might occur in simulated planetary atmospheres, by exposing them to observed AD Leo flare spectra during their impulsive rise, impulsive decay and gradual decay phases, and found that the effect on an atmosphere is more significant during the impulsive phases.

We therefore turned to look for trends between flare peak luminosity and FWHM, for which neither simple nor complex flares showed any significant correlation. The relationship between the two parameters is shown in the right panel of Figure 11, and the associated Pearson correlation coefficients have $p$-values of 0.62 and 0.73 for single and multi-peak flare components, respectively.

Finally, we remark that the flares with the largest duration and peak luminosity are also associated with higher stellar rotation levels and complexity. Figure 12 shows the relationship between these parameters, by colouring the flare *L-d* pairs according to the stellar $v \sin i$ on the left, and the number of flare peaks on the right. Despite the fact that low $v \sin i$ values might be both due to a truly low rotation velocity or to highly inclined stellar rotation axes with respect to the plane of the sky, we qualitatively recovered the same finding as Raetz et al. (2020), who used actual stellar rotation periods as a proxy for stellar activity.

### 6.4. Flare energy

Flare energies were calculated following Davenport (2016), that is, by multiplying the quiescent flare luminosities by the integral under the flares (which is measured in time units). The median energy value is similar between simple and complex flare components: $1.3 \times 10^{30}$ and $3.4 \times 10^{30}$ erg, respectively. However, Figure 13 shows a difference in the high-energy part of the respective distributions: the hypothesis that the two of them belong to the same one is rejected by a KS test with a $p$-value of $\sim 10^{-11}$. This is due to the fact that the complex flare component distribution has a larger number of events in the $10^{31} - 10^{34}$ erg range compared to simple events.

Based on a simple model of the energy *E* released during a magnetic reconnection event, Maehara et al. (2015) suggested that on solar type stars flare duration *d* relates to the event energy as $d \propto E^{1/3}$. This relationship was in agreement with their findings on superflares observed on solar-like stars with Kepler short-cadence data, as well as with Namekata et al. (2017)'s results on solar-like stars white-light flares. Different trends were found by Brasseur et al. (2019)'s measurements on GALEX near-UV, and well as by Pietras et al. (2022) and Yang et al.

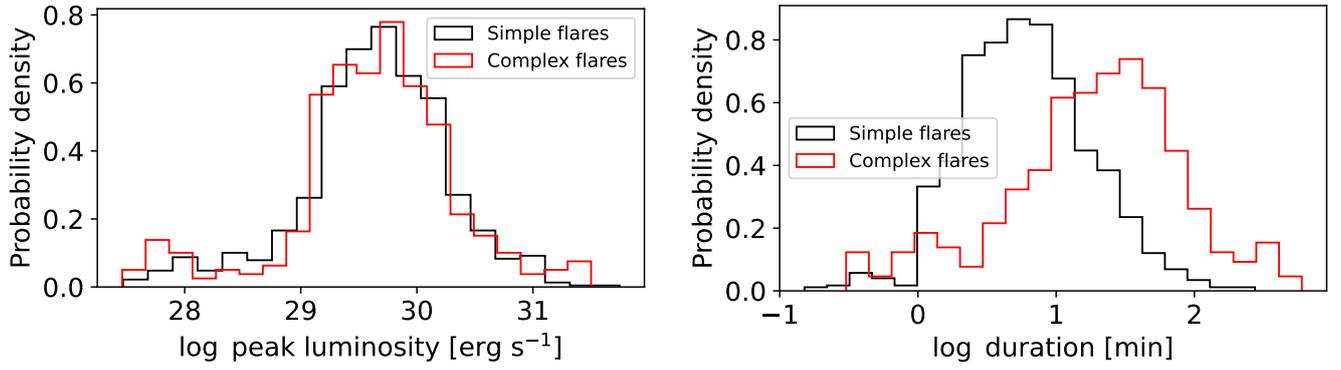

Fig. 10: Distribution of flare luminosity (*left*) and duration (*right*) for single-peaked (*black*) and individual components of multi-peaked (*red*) flares.

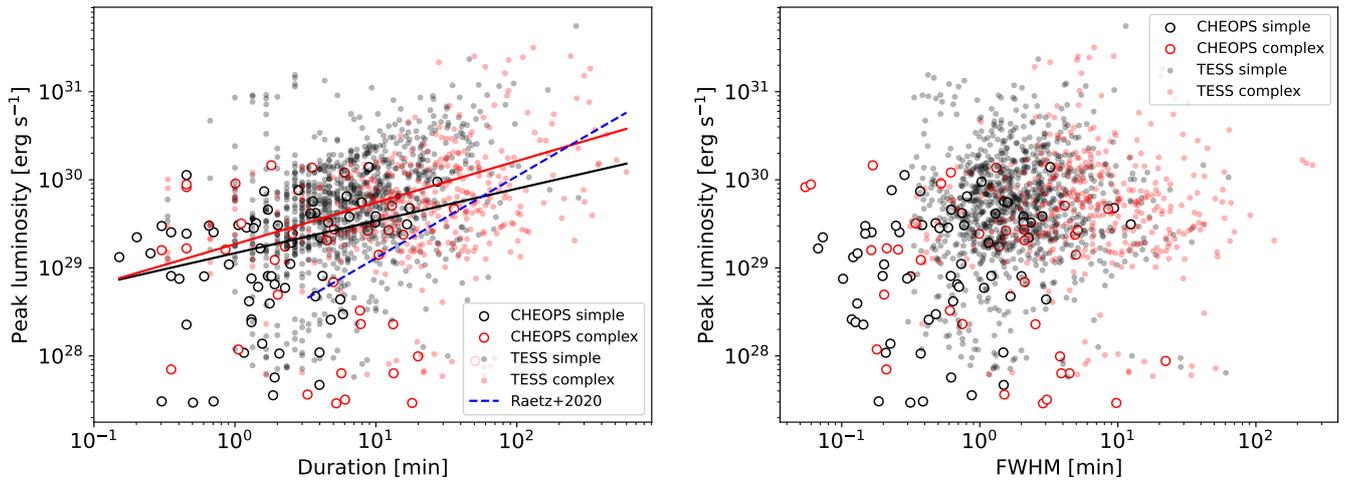

Fig. 11: Flare parameter trends. *Left:* Relationship between flare peak luminosity and duration. Simple and multi-peak flares are separated, as well as the instrument they were detected with. The trend found by Raetz et al. (2020) is shown for the sake of comparison. *Right:* Relationship between flare luminosity and FWHM.

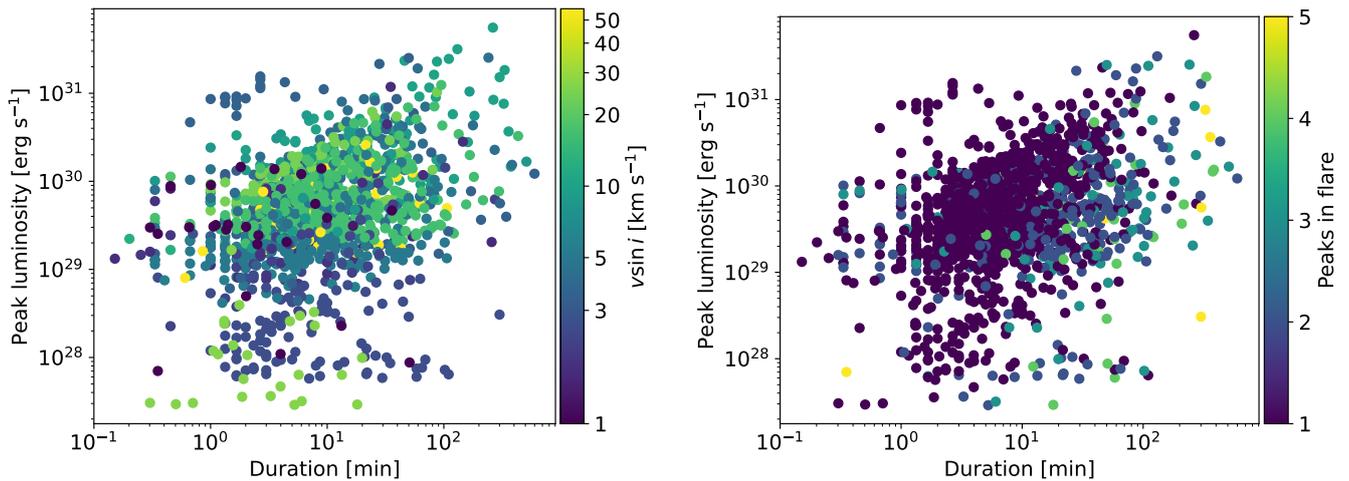

Fig. 12: Flare parameter trends. *Left:* Relationship between flare luminosity and duration, where the stellar $v \sin i$ is represented with the marker colour. *Right:* Same relationship, where the flare complexity is represented with the marker colour. We remind that each point represents a single outburst even in a multi-peak flare, so that the colour refers to the complexity of the event each data point belongs to.

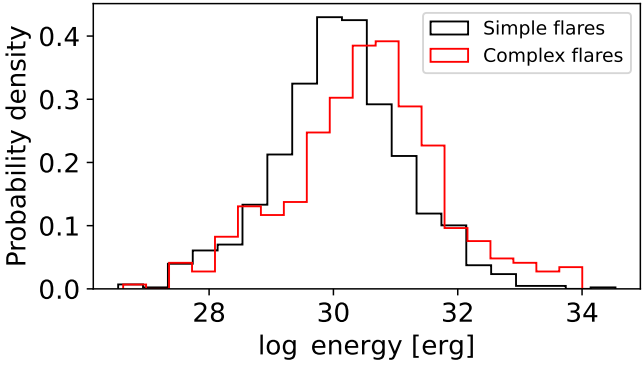

Fig. 13: Distribution of computed flare energies. Single-peaked and individual components of multi-peaked flares are coloured in black and red, respectively.

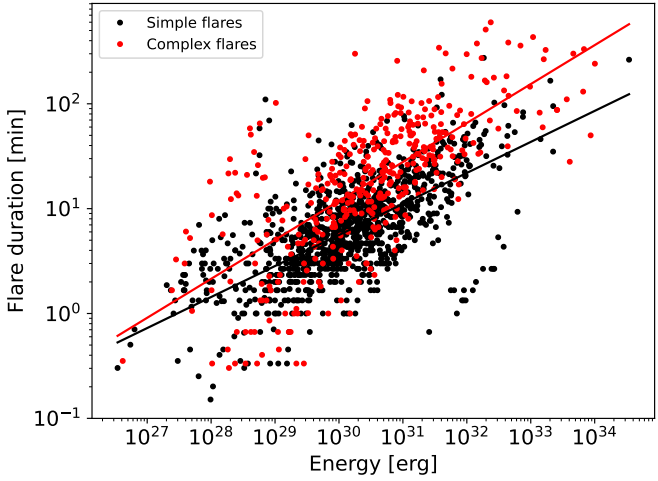

Fig. 14: Flare duration vs energy, divided in simple (black) and complex (red) outburst components.

(2023) on a wider range of spectral types with TESS 2 min data. Deducing from these results a difference in the underlying flare generation process in different types of stars is far from trivial, as it requires budgeting the respective instrumental characteristics. For our dataset, shown in Figure 14, we inspected the flare duration-energy dependence by separating simple and complex flare components, and carrying out a bootstrap fit as for flare luminosity and duration. This yielded

$$\log d = (0.296 \pm 0.004) \log E - (8.1 \pm 0.1) \quad (8)$$

and

$$\log d = (0.37 \pm 0.01) \log E - (10.1 \pm 0.2) \quad (9)$$

for simple and complex flare components, respectively; here, $d$ is measured in min and $E$ in erg. The flares we detected lie in a lower-energy, wider duration region of the parameter space than the one explored by the aforementioned authors (see e.g. Figure 21 in Brasseur et al. 2019), and are broadly in agreement with Maehara et al. (2015)'s model.

As flare impulse $\mathcal{I}$ informs on the most impactful phases of a flare in the surrounding environment, we explored its relationship to flare energy. Even if the integrated energy is comparable at different cadence (Raetz et al. 2020), we statistically observed that the measured impulse tends to increase with time resolution. This is illustrated in Figure 15, where we show the flare impulse and energy for our data set at native (3 and 20 s for CHEOPS and TESS, respectively) and binned to 1-min cadence; when analysing the binned data, we rescaled the retrieved parameters on the basis of injection tests using light curves rebinned at the same cadence.

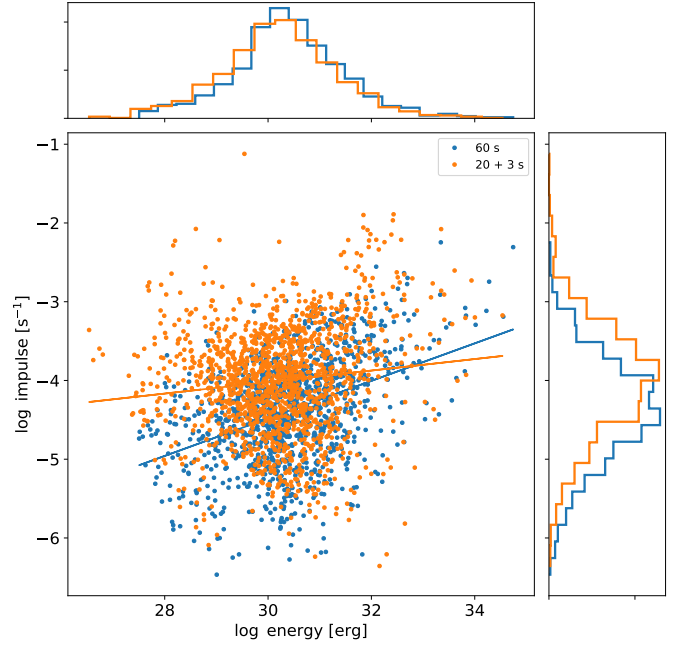

Fig. 15: Relationship between flare impulse and energy at native (orange) and binned to 1-min (blue) cadence. The upper and right panels show the energy and impulse marginalised distributions, respectively.

More in detail, the impulse distribution at native cadence is significantly shifted to higher values with respect to the binned-data distribution: the two of them are distinguished with $p$-value $\sim 10^{-33}$ by a KS test. We also recovered two different trends for the data at native and binned cadence, that is,

$$\log \mathcal{I} = (0.073 \pm 0.004) \log E - (6.2 \pm 0.1) \quad (10)$$

and

$$\log \mathcal{I} = 0.238^{+0.005}_{-0.006} \log E - (11.6 \pm 0.2) \quad (11)$$

for native and binned cadence, respectively, and where impulse is measured in s$^{-1}$. This suggests that the high-time resolution monitoring of stellar hosts might be a relevant factor to correctly estimate the impact of high-rate, low-energy flares onto the evolution of exoplanet atmospheres.

### 6.5. Parameter distributions

The flare energy distribution is generally assumed to follow a power law, that is,

$$N(x)\,dx \propto x^{-\alpha}\,dx, \quad (12)$$

with $\alpha > 0$, in the so-called 'inertial range' $x_1 \leq x \leq x_2$. In this case, $x$ is flare energy. One of the hypotheses to explain this behaviour is that stellar flares are Self-Organised Critical (SOC) systems (Bak et al. 1988; Lu & Hamilton 1991), which are known for their fundamental properties such as spatial extension, duration and delivered energy to be at least partly scale-invariant. Standard SOC system parameters such as peak flux, duration, and energy are distributed according to power laws, but slight deviations might provide indications about the mechanisms that power them (Kunjaya et al. 2011). For example, extreme or 'Dragon-King' events might depart from this relationship (Sornette 2009; Sornette & Ouillon 2012), and a statistical exploration of their distributions might add constraints to investigations based on purely physical principles (Karoff et al. 2016). The detection of irregularities with respect to theoretical power laws and other expected SOC properties is also made possible by the ever-increasing precision in observations (e.g. Sheikh et al. 2016), and might provide insights on mutually triggered, sympathetic flares (Aschwanden 2019). In particular, Dragon-King events appeared to be rare in a sample of hard X ray solar and Kepler stellar flares (Aschwanden 2019). However, complex flare events are not often split into their individual components, which might carry a certain contamination from memory processes (Lei et al. 2020): with the algorithm we developed, we explored whether the contribution from complex outbursts can be distinguished from the one of single-peaked events. While we did not explore the likelihood of complex outbursts as being due to actual sympathetic flares or to independent simultaneous events (e.g. Wheatland 2006; Wheatland & Craig 2006), we consider at least a part of the complex outbursts we observed to be likely sympathetic.

Several factors need to be taken into account when fitting power laws, and one of the most crucial is to estimate the likelihood that other distributions might better represent the data. Following Verbeeck et al. (2019), we chose a log-normal distribution as a plausible alternative, as it might indicate the need for a revision of the basic flare triggering mechanisms. This might be necessary, for example, if not all magnetic energy were released during an outburst (Kunjaya et al. 2011), if its originating magnetic elements went through a fragmentation process (Bogdan et al. 1988), or also if flares were a result of MHD turbulence, as was suggested from the study of flare waiting times (Boffetta et al. 1999; Greco et al. 2009; Norman et al. 2001; Watkins et al. 2016; Lei et al. 2020). Comparing the description of flare parameter distributions can provide additional details to this decades-long-debate.

In the analysis here presented, we neglected the differences between the stars in our sample and considered them to be representative of a 'prototypical' late-K/M star, in order to increase the statistics on the parameter distributions. We fitted the slope $\alpha$ for the flare energy, peak luminosity, and duration distributions following Clauset et al. (2009) and Klaus et al. (2011)'s prescriptions. This method is at least as statistically stable as a fit to a log-log histogram and can provide the likelihood ratio between a power law and another model description of the data. Moreover, it allows a formal determination of $x_1$ (see Equation 12) through minimisation of the distance between the data and the model. This is another key aspect in the estimation of power law parameters, which is often left to subjective evaluation. Adopting similar statistical tools, Verbeeck et al. (2019) found that a log-normal distribution is a preferred description for a sample of about 17000 solar X-ray flares collected between 2010 and 2018.

For our analysis, we used the POWERLAW package (Alstott et al. 2014), where the above described statistical framework is implemented. The top-left panel of Figure 16 shows the complementary cumulative distribution function (CCDF) of the observed flares as a function of their bolometric energy $E$ for flares with only one or multiple resolved components, respectively. In this and the following panels of the same Figure, the CCDF has been scaled to its value at $x_1$. The fitted value for $\alpha$, the normalised likelihood ratio $R$ of the power law against log-normal model, the corresponding $p$-value against the null hypothesis of the descriptions being equivalent and the lower bound of the inertial range used for the fit $x_1$ are reported in Table 2 ($R > 0$ indicates a preference for the power law description, and $R < 0$ the alternative case). The $\alpha$ values for the two distributions are in agreement and are both compatible with most results in the literature, and particularly with the 1.50-1.75 range of flare avalanche models (e.g. Litvinenko 1996). However, we notice the energy distribution of complex events is poorly fitted by a power law, and that a log-normal is a better description with a $p$-value of 0.03. For the single-peak and the combined distribution of single and combined events, the preference for a log-normal is only marginal, so that we can compare them to the predictions of SOC models: in particular, a standard SOC model with $\alpha = 1.44$ (Aschwanden 2022) is in $2\sigma$ agreement with our results for the simple flare distribution, contrarily to Aschwanden (2022)'s three-dimensional fractal energy model. This latter is characterised by $\alpha = 1.80$, which is at $5\sigma$ distance from the result for simple events, but within $2.5\sigma$ from the combined distribution fit.

We repeated the same test for flare peak amplitude and duration, and report our results in the middle and lower panels in the left column of Figure 16. Here, the $\alpha$ slopes predicted by the standard SOC model for peak flux and avalanche duration are $\simeq 1.67$ and 2 (Aschwanden 2022), and are 9 and $4\sigma$ away from our results for the combined distributions, respectively. Overall, we found a $2\sigma$ agreement for the peak luminosity distribution of simple and complex flare components, and a $3.3\sigma$ distance for the respective duration distribution parameters.

The most recent evaluations of $\alpha$ in the flare energy distribution used large samples of 2 min cadence data from the TESS mission, and reported no significant variation of the power law index within earlier and later M stars (Feinstein et al. 2022). The current debate is open, as a tentative $\alpha$ dependence on spectral type was found in other studies (e.g. Yang et al. 2023). This motivated us to inspect the parameter distributions for partially convective (spectral type earlier than M3V) and fully convective (later than M3V) stars, without distinguishing simple and complex flare components. In the right column of Figure 16, we plot the corresponding flare energy, peak luminosity and duration scaled CCDFs. The $\alpha$ indices of the energy distributions are in $\sim 1\sigma$ agreement, while the luminosity and duration distributions are at 4 to $5\sigma$ distance. The energy distributions, moreover, are compatible with the $1.50 - 1.75$ range in both cases, as well as at $2\sigma$ agreement with the three-dimensional fractal energy model. In this analysis, selection effects cannot be excluded, as the flare S/N, which peaks at short wavelengths, is expected to increase for redder and cooler stars. For these latter, the detection of low-energy events might be easier, given the same photometric noise level.

In terms of the comparison between power-law and log-normal descriptions of the parameter distributions, we found variations depending on the parameter and subset, but in all cases only marginal. To inspect whether a larger sample could increase the statistical significance of one model description compared to the other, for each outlined case we used POWERLAW to simulate data sets corresponding to the fitted power law parameters, each with $10^4$ samples. For each of these data sets, we fitted an other power law and log-normal distribution, and compared the resulting likelihood ratio. We found that the duration distribu-tions could be better constrained with a larger sample, a fact that could be addressed by further dedicated analyses.

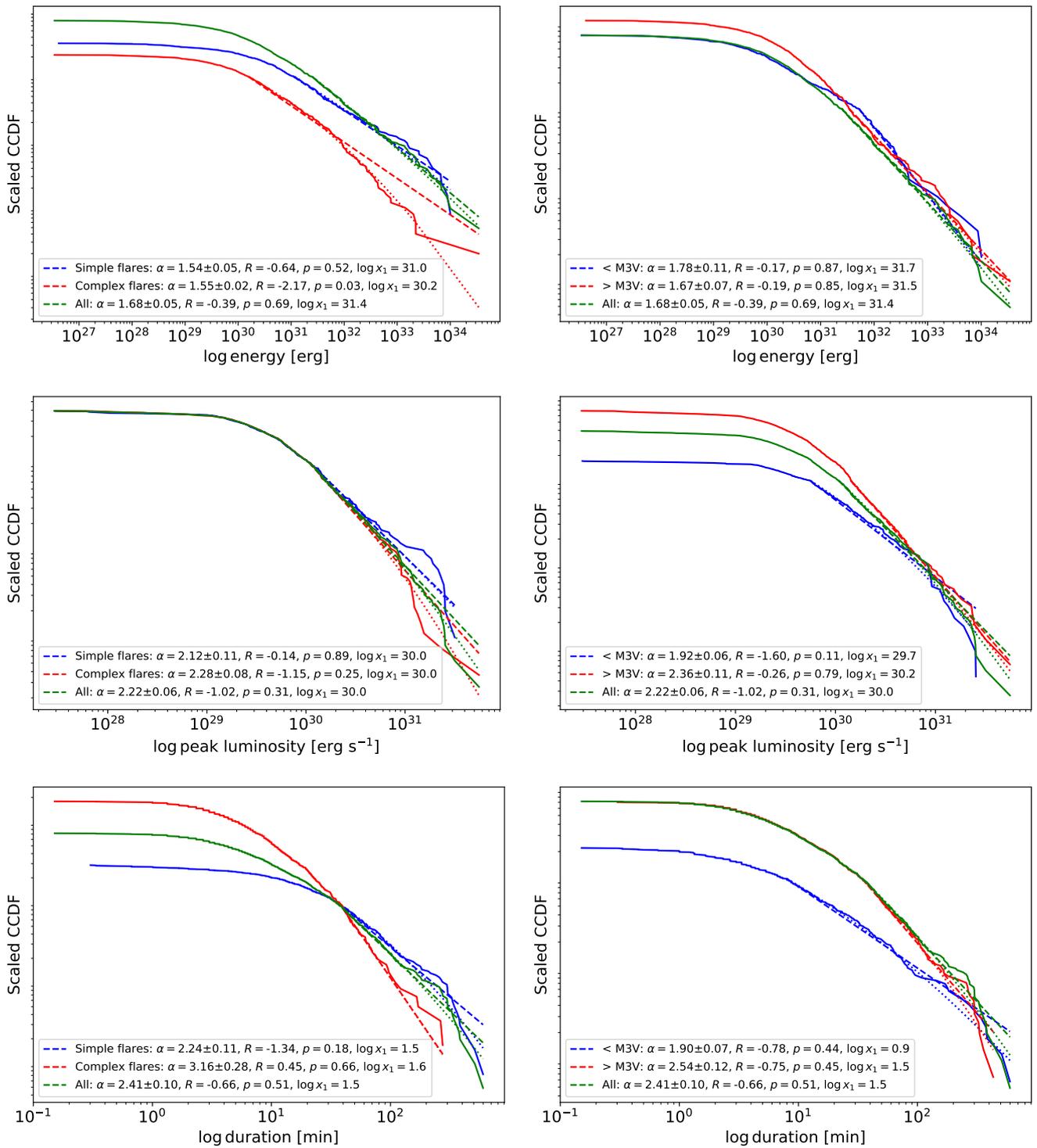

Fig. 16: Complementary cumulative distribution functions (CCDFs) for flare parameters. *Left column:* scaled complementary cumulative distribution functions (CCDF) for flare energy (*top*), peak luminosity (*centre*) and duration (*bottom*), separated between simple flares (blue), complex flare components (red), and the full sample (green). Observed distributions, power-law fits, and log-normal distribution fits are represented with lines, dashed lines, and dotted lines, respectively. *Right column:* same as the left column, but where the flare sample is divided in events occurring on stars earlier (blue) or later (red) than M3V. In each panel, the legend indicates the fitted power-law coefficient $\alpha$, the normalised likelihood ratio $R$ between the power-law and log-normal fits, its associated $p$-value, and the lower bound of the inertial range used for the fit, $x_1$.

|  | $\alpha$ | $R$ | $p$-value | $x_1$ [log erg] |
|---|---|---|---|---|
| *Simple flares* | | | | |
| Energy | 1.54 ± 0.05 | −0.64 | 0.52 | 31.0 |
| Peak luminosity | 2.11 ± 0.11 | −0.14 | 0.89 | 30.0 |
| Duration | 2.24 ± 0.11 | −1.34 | 0.18 | 1.5 |
| *Complex flare components* | | | | |
| Energy | 1.55 ± 0.02 | −2.17 | 0.03 | 30.2 |
| Peak luminosity | 2.28 ± 0.08 | −1.15 | 0.25 | 30.0 |
| Duration | 3.16 ± 0.28 | −0.01 | 1.00 | 1.6 |
| *Partially convective stars* | | | | |
| Energy | 1.78 ± 0.11 | −0.17 | 0.87 | 31.7 |
| Peak luminosity | 1.92 ± 0.06 | −1.60 | 0.11 | 29.7 |
| Duration | 1.90 ± 0.07 | −1.18 | 0.24 | 0.9 |
| *Fully convective stars* | | | | |
| Energy | 1.67 ± 0.07 | −0.19 | 0.85 | 31.5 |
| Peak luminosity | 2.36 ± 0.11 | −0.26 | 0.79 | 30.2 |
| Duration | 2.54 ± 0.12 | −0.84 | 0.40 | 1.5 |
| *All stars* | | | | |
| Energy | 1.68 ± 0.05 | −0.39 | 0.69 | 31.4 |
| Peak luminosity | 2.22 ± 0.06 | −1.02 | 0.31 | 30.0 |
| Duration | 2.41 ± 0.10 | −0.91 | 0.35 | 1.5 |

Table 2: Fitted power-law coefficients $\alpha$, normalised likelihood ratio $R$, corresponding $p$-value, and lower bound for the inertial range $x_1$ for the cumulative distribution of each flare parameter and subset.

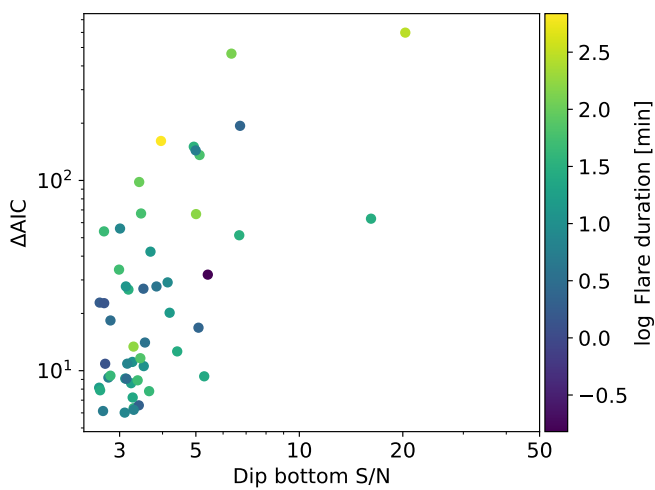

Fig. 17: Pre-flare dip AIC between a model including a pre-flare dip and a model without it as a function of the candidate dip S/N. Only dips with the criteria outlined in Section 6.6 are shown.

### 6.6. Pre-flare dips

To inspect for pre-flare dip candidates, we filtered our dip fits and selected only those with 1) $\Delta$AIC> 6 with respect to the model without dip, 2) a S/N> 2.5 for the dip amplitude with respect to the correlated noise level in a time window as wide as the fitted dip, 3) an overall flare reduced $\chi^2 < 10$, and 4) a dip duration of at least three data points (measured as the sum of the dip width on both sides). In Figure 17, we display the dip candidates that passed these criteria, and colour the points according to the duration of the associated flare (or the total duration of the outburst in case of a complex flare). We report this information because, by visual inspection, we noticed that in most cases the dip is associated with long-duration flares, where it is likely that the light curve smoothing process has artificially reduced the flux before the rise phase of the flare, creating a false dip effect.

After a visual inspection, we only retained one event as valid dip candidate, which is illustrated in Figure 18. It was observed with CHEOPS on V1054 Oph: interestingly, a multi-band pre-flare dip was reported by Ventura et al. (1995) on the same star. The relative amplitude of the candidate, which occurred $\simeq 2.7$ min before the beginning of the corresponding flare, is $(0.17 \pm 0.02)\%$, and its width $(47 \pm 13)$ s. Its following flare was also very little energetic, with $E \simeq 6.6 \times 10^{28}$ erg: this is one of the smallest energies we measured. Comparatively, the one found by Ventura et al. (1995) right before a flare event had a relative amplitude of $(6 \pm 3)\%$ in the $V$ band, and an exceptional duration of $\simeq 36$ min. Its following flare could not be observed in the $V$ band, but presented $E = 1.66 \times 10^{32}$ erg in the $B$ band. Its lack of visibility in the $V$ band might make it comparable to the very small-energy flare we detected.

All in all, no information that could have helped us validate our candidate was available in other wavelengths. Therefore, we consider our detection only tentative, and refrained from a more detailed analysis.

### 6.7. Quasi-periodic pulsations

QPPs can be empirically distinguished from multi-peak flares by the quasi-periodicity of the flux rises and decays. We visually searched for QPP candidates in our validated complex flares: as our algorithm does not model QPPs, it is likely that some of them are misinterpreted as multi-peak flares. After rejecting those with the most irregular sequences of sharp peaks and smooth 'bumps', we isolated one and 14 QPP candidates in the CHEOPS and TESS light curves, respectively: this is about 1% of the full sample, close to recent reports (Balona & Abedigamba 2016; Pugh et al. 2016). One of our candidate QPPs is shown in Figure 19: in the left panel, we show the highest-likelihood multi-peak flare model, which we used to extract the mean periodicity of the oscillation as the average distance between consecutive flare peaks. On the right panel, we show a model with a single, smooth-peak flare shape fitted on the complex flare profile: from the half-difference between the largest and smallest flux values we extracted the oscillation amplitude. Following Howard & MacGregor (2022), we examined the ratio between this quantity and the amplitude of the single-peak flare used to derive the flux residuals. The other candidates, on which we applied the same method, are reported in Appendix B. The 'waiting times' we determined span the 4-50 min range, in agreement with previous results (e.g. Pugh et al. 2016; Vida et al. 2019; Ramsay et al. 2021; Million et al. 2021; Howard & MacGregor 2022).

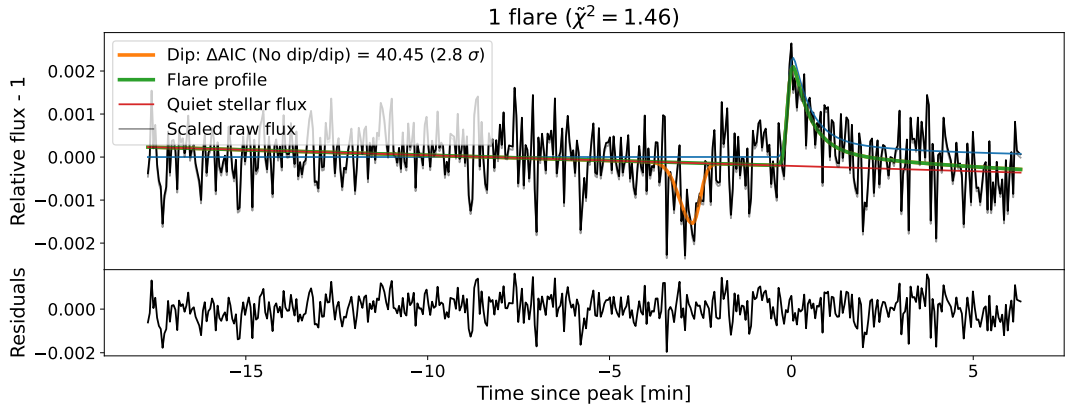

Fig. 18: Pre-flare dip tentative detection on V1054 Oph.

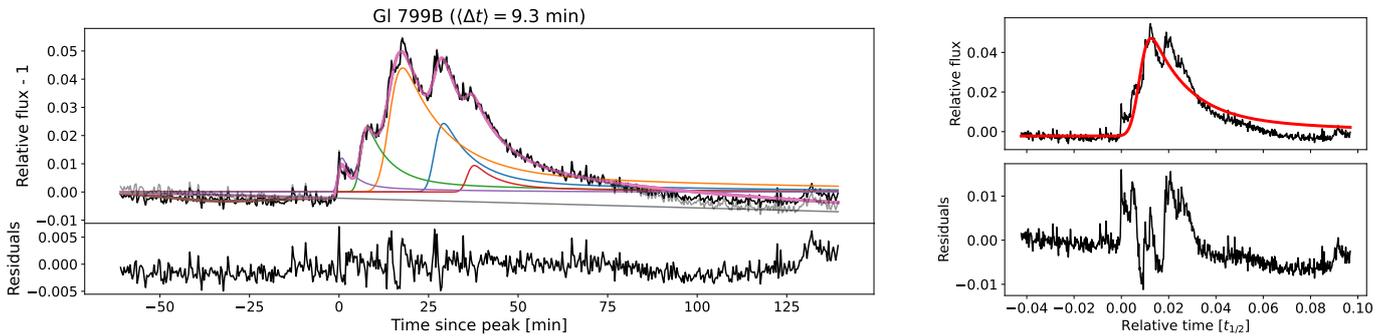

Fig. 19: QPP candidate on Gl 799 B. *Left:* the mean time between consecutive peaks is reported in the title. This QPP candidate is probably the same one reported by Howard & MacGregor (2022) (their Figure 7, left panel), for which we recover a similar period to their $\simeq 7.5$ min. The individual flare components that were identified by our algorithm are drawn with different colours, and the total model is represented with a thick line. *Right:* on the top sub-panel, fit to the flare profile with a single-peak model. On the lower sub-panel, the residuals used to estimate the oscillation amplitude.

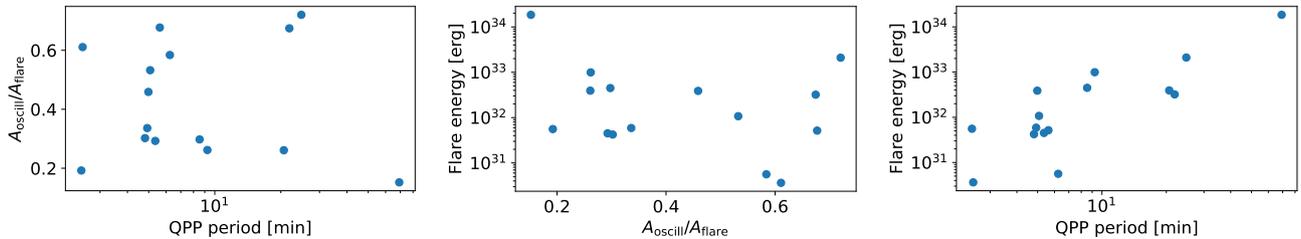

Fig. 20: Relationships between estimated QPP parameters.

In Figure 20, we plot the QPP parameters we derived as in Howard & MacGregor (2022), and recovered similar tentative correlations: in particular, a negative trend between flare energy (i.e. the single-smoothed flare fitted to the complex profile) and oscillation amplitude, and a positive trend between flare energy with the QPP duration.

To be conservative, the flare sample just presented was not included when deriving flare parameters in the previous parts of this study (even if we verified that it does not significantly affect any result). Additionally, we assessed the possible contamination of undetected QPPs in the fits of all the other flare profiles. To do this, we first examined the periodicities with False Alarm Probability (FAP) < 1% that emerged from the Lomb-Scargle periodograms (LSP) of the flare fit residuals. The calculation was carried out using Baluev (2008)'s method, implemented in the `astropy` tools we used: 18% of the flare residuals present significant peaks, and their histogram is shown in Figure 21. The median peak of the distribution corresponds to $\simeq 67$ min, which is larger than the duration of most complex flare components (right panel of Figure 10), time delay between consecutive flares (Figure 9), and reported values in the literature for QPP periods, which span the $\sim 2 - 72$ min range (e.g. Pugh et al. 2016; Vida et al. 2019; Ramsay et al. 2021; Million et al. 2021; Howard & MacGregor 2022); from this calculation, we excluded a candidate found on AD Leo, highlighted in Appendix B, because of its corresponding poor fit. This suggests that the correlated signal present in some residual light curves is not related to QPPs, but to uncorrected systematics or poor fits; we recall that we consider unlikely for stellar granulation to be detectable in our light curves, given their S/N and the granulation expected signal.

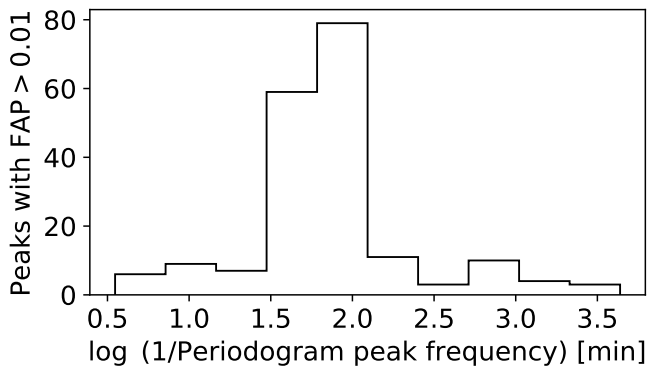

Fig. 21: Histogram of the flare residuals periodicity corresponding to LSP peeks with FAP< 1%.

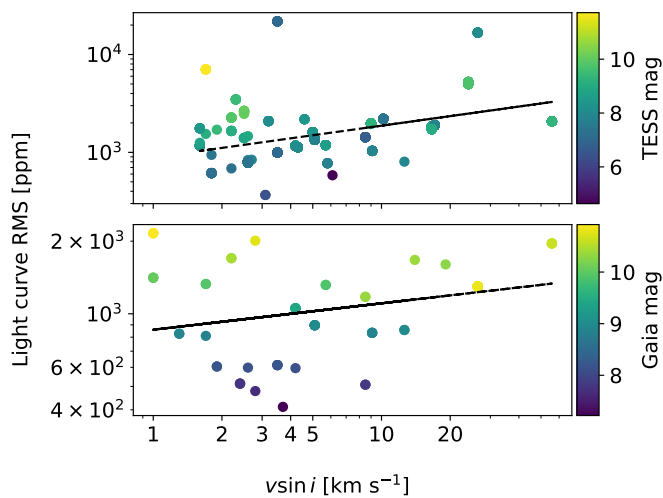

Fig. 22: Correlation between flattened light curve scatter ($y$-axis) and stellar $v \sin i$ ($x$-axis) for the stars for which this quantity was available. TESS and CHEOPS are represented in the top and bottom panel, respectively. A log-log fit is shown with a dashed line.

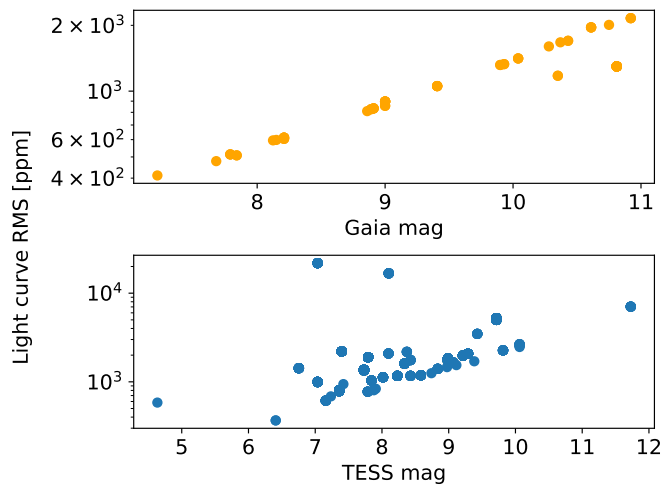

Fig. 23: Correlation between flattened light curve scatter ($y$-axis) and stellar magnitude. CHEOPS observations with Gaia magnitudes and TESS light curves with TESS magnitudes are represented in the top and bottom panel, respectively.

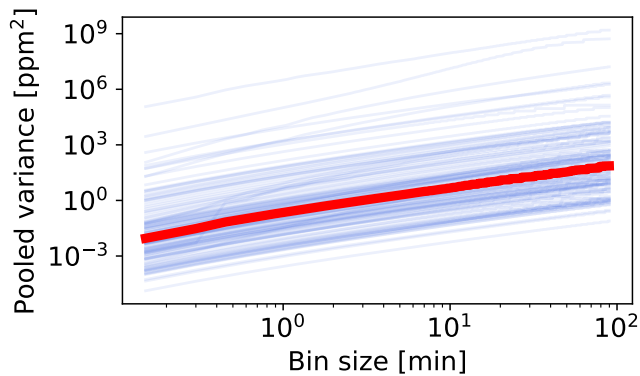

Fig. 24: Pooled variance of the detrended CHEOPS light curves as a function of bin size, shown with blue lines. The median pooled variance curve is shown with a thick red line.

### 6.8. Undetected flares and light curve scatter

Stelzer et al. (2016) reported a correlation between photometric scatter in the detrended light curves and stellar rotation period, and argued that this might be a signature of undetected starspots or flares. As shown in Figure 22, we observed a similar relationship for the targets for which we have a $v \sin i$ measurement. The Pearson correlation coefficient between the two quantities excludes the no-correlation hypothesis with a $p$-value of $\sim 10^{-19}$ and $10^{-5}$ for TESS and CHEOPS data, respectively. The positive correlation in TESS light curves is mainly driven by the points with RMS > 5000 ppm, which correspond to GJ 65, GJ 3304, G 214-14 and AD Leo, which we inspected to exclude any anomaly.

However, contrarily to Stelzer et al. (2016), we also noticed a very strong correlation between flattened light curve RMS and stellar magnitude, as shown in Figure 23. This makes the impact of stellar activity features in the photometric scatter unlikely in our case.

We remark, moreover, that a pooled variance analysis (Dobson et al. 1990; Donahue & Baliunas 1992; Donahue 1993; Lanza et al. 2003) of the flattened CHEOPS light curves does not seem to approach any 'basal level' in the light curves scatter. This might be related, for example, to an approaching 'flare background', as suggested by Stelzer et al. (2016), and would correspond to a flattening in the pooled variance at the lowest cadences. The pooled variance of our CHEOPS observations is shown in Figure 24.

# 7. Discussion and conclusions

The use of high-cadence, high precision photometry from CHEOPS and TESS allowed us to probe unexplored regions of the white-light flare parameter space in late-K and M stars. In this study, moreover, we adopted the same analytical model for the profile of single-peak and multiple-peak outbursts, and searched for differences in the distributions of the respective flare luminosity, duration, and energy. Using this, we also investigated whether power laws are the most likely description for simple and complex flare parameter distributions, as most studies generally assumed, but which has recently been challenged in studies of solar flares. Finally, we inspected for the presence of pre-flare dips and MHD-driven QPPs, and found no indication of an undetected flare background below the data noise level.

We confirmed the results by Hawley et al. (2014), who found that complex flares can reach longer durations than simple events, even after the individual flare components are resolved. We did not retrieve significant differences between flare parameter correlations for events lasting more than a few minutes (already thoroughly investigated with Kepler/K2 and 2-min TESS data in previous works) and shorter ones. However, our analysis suggests higher flare impulses can be reached on cool stars, compared to previous findings. This indicates that high-cadence photometric monitoring of planet hosts might be crucial to correctly estimate the high-energy, time-limited flux exoplanet atmospheres might be exposed to. At lower cadence, the high-impulse stellar irradiation might be missed, and the only flare energy distribution might provide incomplete information to compute trustworthy models for planetary atmospheres. The amount of deposited energy is needed to compute, for example, variations in ozone abundance and rates of water photolysis (e.g. Segura et al. 2010; Loyd et al. 2018). This result is particularly relevant in the context of potentially habitable planet studies.

We retrieved ∼ 40% and ∼ 30% percentage of complex flares in both CHEOPS and TESS data, in agreement with previous studies concentrating on M stars (e.g. Davenport et al. 2014; Howard & MacGregor 2022). This suggests that stellar flares observed with future facilities will require the modelling of complex flares to correctly unveil their properties. For example, PLATO (Rauer et al. 2014) will monitor its targets at 25 s cadence. Its time resolution will reach 2.5 s for bright stars, which will also be observed in blue and red filters: therefore, we can expect it to provide large statistics about flares happening on several stellar types during its years-long continuous surveys. Ariel (Tinetti et al. 2018), instead, will be capable of 1 s cadence measurements, and might happen to resolve flare fine details during some of its exoatmosphere recognition campaigns. Both missions would benefit from further investigation in the degeneracies involved in complex flare profile fitting and false positive assessment to maximise the scientific return of their data.

Our results are qualitatively consistent with the part of Tovmassian et al. (2003)'s model that posits the peak-bump flare profile is the fundamental flare shape. Instead, we could not validate the part of their model which considers a starspot-induced, pre-flare dip as prevalent in flare profiles. Reported dip observations were made in the $U$, $B$, $V$, $R$ and $I$ bands, and spectroscopically in $H_\alpha$ (Leitzinger et al. 2014, and references therein): their signal would likely fall below the noise level for TESS, while they should be detectable with CHEOPS if they happened on a regular basis. However, we tentatively detected only one, $\lesssim$ 1 min pre-flare dip candidate, and through injection tests we found that dips are likely to be missed with CHEOPS because of the combination of data gaps and residual instrument systematics.

Our algorithm does not include the modelling of QPPs, which we visually searched in the data. Overall, ∼ 1% of our validated flare sample could be classified as QPP candidates. We measured amplitude and period of such possible oscillations and found results in agreement with the literature, but could only identify tentative properties because of lack of statistics.

We dedicated a particular focus to the analysis of flare energy, peak luminosity, and duration distributions, and did not find significant differences in the energy and peak luminosity distributions of simple and complex flare components, contrarily to their duration distributions. As expected, we could not reach energies that might inform us on the role of micro- and nanoflares in coronal heating. Moreover, we evaluated the likelihood that an alternative description to power laws might better represent flare parameter distributions, but only found marginal preference for log-normal models with respect to power-laws in most cases, with the exception of the energy distribution of complex flare components. Verbeeck et al. (2019) found significant indications that log-normal distributions are a better fit to solar data: in this case, the unavoidable lower S/N of our data might be a limitation, and requires further investigation. In any case, our analysis supports the importance of a formal estimation of the inertial range used for the fit of power law parameters.

We also separated the flare parameter distributions occurring in partly and fully convective stars. We found significant differences in the slope of the power laws associated to earlier and later M stars peak luminosity and duration, which might indicate a selection effect as well as a transition in the generation of the magnetic field from the $\alpha\Omega$ to the $\alpha^2$ dynamo process (e.g. Chabrier & Küker 2006).

The detailed exploration of power law distributions and their alternatives needs large samples to be statistically stable. Statistical frameworks such as the one by Clauset et al. (2009) and (Klaus et al. 2011) were built to maintain stability also in the case of samples with just a few hundred elements; however, we showed through simulated data sets that investigations on larger and possibly more diverse samples could shed more light on these aspects. This will surely benefit from the ultra-high precision, high-cadence, and in few cases two-coloured PLATO observations.

*Acknowledgements.* We thank Antonino F. Lanza and Daniel Brito de Freitas for their suggestions and the very fruitful discussions we had. This research has made use of the SIMBAD database, operated at CDS, Strasbourg, France. Our search was helped by the use of the ASTROQUERY package (Ginsburg et al. 2019). CHEOPS is an ESA mission in partnership with Switzerland with important contributions to the payload and the ground segment from Austria, Belgium, France, Germany, Hungary, Italy, Portugal, Spain, Sweden, and the United Kingdom. The CHEOPS Consortium would like to gratefully acknowledge the support received by all the agencies, offices, universities, and industries involved. Their flexibility and willingness to explore new approaches were essential to the success of this mission. CHEOPS data analysed in this article will be made available in the CHEOPS mission archive (https://cheops.unige.ch/archive_browser/). LBo, GBr, VNa, IPa, GPi, RRa, and GSc acknowledge support from CHEOPS ASI-INAF agreement n. 2019-29-HH.0. ABr was supported by the SNSA. PM acknowledges support from STFC research grant number ST/R000638/1. S.G.S. acknowledge support from FCT through FCT contract nr. CEECIND/00826/2018 and POPH/FSE (EC). The Portuguese team thanks the Portuguese Space Agency for the provision of financial support in the framework of the PRODEX Programme of the European Space Agency (ESA) under contract number 4000142255. V.V.G. is an F.R.S-FNRS Research Associate. ZG was supported by the VEGA grant of the Slovak Academy of Sciences No. 2/0031/22 and by the Slovak Research and Development Agency - the contract No. APVV-20-0148. GyMSz acknowledges the support of the Hungarian National Research, Development and Innovation Office (NKFIH) grant K-125015, a a PRODEX Experiment Agreement No. 4000137122, the Lendulet LP2018-7/2021 grant of the Hungarian Academy of Science and the support of the city of Szombathely. DG gratefully acknowledges financial support from the CRT foundation under Grant No. 2018.2323 "Gaseousor rocky? Unveiling the nature of small worlds". YAl acknowledges support from the Swiss National Science Foundation (SNSF) under grant 200020_192038. We acknowledge financial support from the Agencia Estatal de Investigación of the Ministerio de Ciencia e Innovación MCIN/AEI/10.13039/501100011033 and the ERDF "A way of making Europe" through projects PID2019-107061GB-C61, PID2019-107061GB-C66, PID2021-125627OB-C31, and PID2021-125627OB-C32, from the Centre of Excellence "Severo Ochoa" award to the Instituto de Astrofísica de Canarias (CEX2019-000920-S), from the Centre of Excellence "María de Maeztu" award to the Institut de Ciències de l'Espai (CEX2020-001058-M), and from the Generalitat de Catalunya/CERCA programme. We acknowledge financial support from the Agencia Estatal de Investigación of the Ministerio de Ciencia e Innovación MCIN/AEI/10.13039/501100011033 and the ERDF "A way of making Europe" through projects PID2019-107061GB-C61, PID2019-107061GB-C66, PID2021-125627OB-C31, and PID2021-125627OB-C32, from the Centre of Excellence "Severo Ochoa" award to the Instituto de Astrofísica de Canarias (CEX2019-000920-S), from the Centre of Excellence "María de Maeztu" award to the Institut de Ciències de l'Espai (CEX2020-001058-M), and from the Generalitat de Catalunya/CERCA programme. S.C.C.B. acknowledges support from FCT through FCT contracts nr. IF/01312/2014/CP1215/CT0004. C.B. acknowledges support from the Swiss Space Office through the ESA PRODEX program. This work has been carried out within the framework of the NCCR PlanetS supported by the Swiss National Science Foundation under grants 51NF40_182901 and 51NF40_205606. ACC acknowledges support from STFC consolidated grant numbers ST/R000824/1 and ST/V000861/1, and UKSA grant number ST/R003203/1. P.E.C. is funded by the Austrian Science Fund (FWF) Erwin Schroedinger Fellowship, program J4595-N. This project was supported by the CNES. S.S. acknowledges support from CNES, the Programme National de Planétologie (PNP), and the Programme National de Physique Stellaire (PNPS) of CNRS-INSU. The Belgian participation to CHEOPS has been supported by the Belgian Federal Science Policy Office (BELSPO) in the framework of the PRODEX Program, and by the University of Liège through an ARC grant for Concerted Research Actions financed by the Wallonia-Brussels Federation. L.D. thanks the Belgian Federal Science Policy Office (BELSPO) for the provision of financial support in the framework of the PRODEX Programme of the European Space Agency (ESA) under contract number 4000142531. This work was supported by FCT - Fundação para a Ciência e a Tecnologia through national funds and by FEDER through COMPETE2020 through the research grants UIDB/04434/2020, UIDP/04434/2020, 2022.06962.PTDC. O.D.S.D. is supported in the form of work contract (DL 57/2016/CP1364/CT0004) funded by national funds through FCT. B.-O. D. acknowledges support from the Swiss State Secretariat for Education, Research and Innovation (SERI) under contract number MB22.00046. This project has received funding from the Swiss National Science Foundation for project 200021_200726. It has also been carried out within the framework of the National Centre of Competence in Research PlanetS supported by the Swiss National Science Foundation under grant 51NF40_205606. The authors acknowledge the financial support of the SNSF. MF and CMP gratefully acknowledge the support of the Swedish National Space Agency (DNR 65/19, 174/18). M.G. is an F.R.S.-FNRS Senior Research Associate. MNG is the ESA CHEOPS Project Scientist and Mission Representative, and as such also responsible for the Guest Observers (GO) Programme. MNG does not relay proprietary information between the GO and Guaranteed Time Observation (GTO) Programmes, and does not decide on the definition and target selection of the GTO Programme. CHe acknowledges support from the European Union H2020-MSCA-ITN-2019 under Grant Agreement no. 860470 (CHAMELEON). SH gratefully acknowledges CNES funding through the grant 837319. KGI is the ESA CHEOPS Project Scientist and is responsible for the ESA CHEOPS Guest Observers Programme. She does not participate in, or contribute to, the definition of the Guaranteed Time Programme of the CHEOPS mission through which observations described in this paper have been taken, nor to any aspect of target selection for the programme. K.W.F.L. was supported by Deutsche Forschungsgemeinschaft grants RA714/14-1 within the DFG Schwerpunkt SPP 1992, Exploring the Diversity of Extrasolar Planets. This work was granted access to the HPC resources of MesoPSL financed by the Region Ile de France and the project Equip@Meso (reference ANR-10-EQPX-29-01) of the programme Investissements d'Avenir supervised by the Agence Nationale pour la Recherche. ML acknowledges support of the Swiss National Science Foundation under grant number PCEFP2_194576. This work was also partially supported by a grant from the Simons Foundation (PI Queloz, grant number 327127). NCSa acknowledges funding by the European Union (ERC, FIERCE, 101052347). Views and opinions expressed are however those of the author(s) only and do not necessarily reflect those of the European Union or the European Research Council. Neither the European Union nor the granting authority can be held responsible for them. A. S. acknowledges support from the Swiss Space Office through the ESA PRODEX program. JV acknowledges support from the Swiss National Science Foundation (SNSF) under grant PZ00P2_208945. NAW acknowledges UKSA grant ST/R004838/1. TWi acknowledges support from the UKSA and the University of Warwick.

[1] INAF, Osservatorio Astrofisico di Catania, Via S. Sofia 78, 95123 Catania, Italy
[2] Department of Astronomy, Stockholm University, AlbaNova University Center, 10691 Stockholm, Sweden
[3] Astrophysics Group, Lennard Jones Building, Keele University, Staffordshire, ST5 5BG, United Kingdom
[4] Weltraumforschung und Planetologie, Physikalisches Institut, University of Bern, Gesellschaftsstrasse 6, 3012 Bern, Switzerland
[5] Center for Space and Habitability, University of Bern, Gesellschaftsstrasse 6, 3012 Bern, Switzerland
[6] Instituto de Astrofisica e Ciencias do Espaco, Universidade do Porto, CAUP, Rua das Estrelas, 4150-762 Porto, Portugal
[7] Aix Marseille Univ, CNRS, CNES, LAM, 38 rue Frédéric Joliot-Curie, 13388 Marseille, France
[8] Space sciences, Technologies and Astrophysics Research (STAR) Institute, Université de Liège, Allée du 6 Août 19C, 4000 Liège, Belgium
[9] HUN-REN-ELTE Exoplanet Research Group, 9700 Szombathely, Szent Imre, h. u. 112, Hungary
[10] ELTE Gothard Astrophysical Observatory, 9700 Szombathely, Szent Imre, h. u. 112, Hungary
[11] Astronomical Institute, Slovak Academy of Sciences, 05960 Tatranská Lomnica, Slovakia
[12] Konkoly Observatory, HUN-REN Research Centre for Astronomy and Earth Sciences, Konkoly Thege út 15-17., H-1121, Budapest, Hungary
[13] CSFK, MTA Centre of Excellence, Budapest, Konkoly Thege út 15-17., H-1121, Hungary
[14] Dipartimento di Fisica, Università degli Studi di Torino, via Pietro Giuria 1, I-10125, Torino, Italy
[15] Instituto de Astrofísica de Canarias, Vía Láctea s/n, 38200 La Laguna, Tenerife, Spain
[16] Departamento de Astrofísica, Universidad de La Laguna, Astrofísico Francisco Sanchez s/n, 38206 La Laguna, Tenerife, Spain

[17] Admatis, 5. Kandó Kálmán Street, 3534 Miskolc, Hungary
[18] Depto. de Astrofísica, Centro de Astrobiología (CSIC-INTA), ESAC campus, 28692 Villanueva de la Cañada (Madrid), Spain
[19] Departamento de Fisica e Astronomia, Faculdade de Ciencias, Universidade do Porto, Rua do Campo Alegre, 4169-007 Porto, Portugal
[20] Space Research Institute, Austrian Academy of Sciences, Schmiedlstrasse 6, A-8042 Graz, Austria
[21] Observatoire astronomique de l'Université de Genève, Chemin Pegasi 51, 1290 Versoix, Switzerland
[22] INAF, Osservatorio Astronomico di Padova, Vicolo dell'Osservatorio 5, 35122 Padova, Italy
[23] Centre for Exoplanet Science, SUPA School of Physics and Astronomy, University of St Andrews, North Haugh, St Andrews KY16 9SS, UK
[24] Institute of Planetary Research, German Aerospace Center (DLR), Rutherfordstrasse 2, 12489 Berlin, Germany
[25] INAF, Osservatorio Astrofisico di Torino, Via Osservatorio, 20, I-10025 Pino Torinese To, Italy
[26] Centre for Mathematical Sciences, Lund University, Box 118, 221 00 Lund, Sweden
[27] Astrobiology Research Unit, Université de Liège, Allée du 6 Août 19C, B-4000 Liège, Belgium
[28] Institute of Astronomy, KU Leuven, Celestijnenlaan 200D, 3001 Leuven, Belgium
[29] Centre Vie dans l'Univers, Faculté des sciences, Université de Genève, Quai Ernest-Ansermet 30, 1211 Genève 4, Switzerland
[30] Leiden Observatory, University of Leiden, PO Box 9513, 2300 RA Leiden, The Netherlands
[31] Department of Space, Earth and Environment, Chalmers University of Technology, Onsala Space Observatory, 439 92 Onsala, Sweden
[32] Department of Astrophysics, University of Vienna, Türkenschanzstrasse 17, 1180 Vienna, Austria
[33] European Space Agency (ESA), European Space Research and Technology Centre (ESTEC), Keplerlaan 1, 2201 AZ Noordwijk, The Netherlands
[34] Institute for Theoretical Physics and Computational Physics, Graz University of Technology, Petersgasse 16, 8010 Graz, Austria
[35] Konkoly Observatory, Research Centre for Astronomy and Earth Sciences, 1121 Budapest, Konkoly Thege Miklós út 15-17, Hungary
[36] ELTE Institute of Physics, Pázmány Péter sétány 1/A, 1117 Budapest, Hungary
[37] IMCCE, UMR8028 CNRS, Observatoire de Paris, PSL Univ., Sorbonne Univ., 77 av. Denfert-Rochereau, 75014 Paris, France
[38] Institut d'astrophysique de Paris, UMR7095 CNRS, Université Pierre & Marie Curie, 98bis blvd. Arago, 75014 Paris, France
[39] Institute of Optical Sensor Systems, German Aerospace Center (DLR), Rutherfordstrasse 2, 12489 Berlin, Germany
[40] Dipartimento di Fisica e Astronomia "Galileo Galilei", Università degli Studi di Padova, Vicolo dell'Osservatorio 3, 35122 Padova, Italy
[41] Department of Physics, University of Warwick, Gibbet Hill Road, Coventry CV4 7AL, United Kingdom
[42] ETH Zurich, Department of Physics, Wolfgang-Pauli-Strasse 2, CH-8093 Zurich, Switzerland
[43] Cavendish Laboratory, JJ Thomson Avenue, Cambridge CB3 0HE, UK
[44] Institut fuer Geologische Wissenschaften, Freie Universitaet Berlin, Maltheserstrasse 74-100,12249 Berlin, Germany
[45] Institut de Ciencies de l'Espai (ICE, CSIC), Campus UAB, Can Magrans s/n, 08193 Bellaterra, Spain
[46] Institut d'Estudis Espacials de Catalunya (IEEC), Gran Capità 2-4, 08034 Barcelona, Spain
[47] Institute of Astronomy, University of Cambridge, Madingley Road, Cambridge, CB3 0HA, United Kingdom

# Appendix A: Target list and TESS proposal reference

Table A.1: List of targets and related parameters.

| Name | GAIA $G$ mag | $T_{\rm eff}$ [K] | $v \sin i$ [km s$^{-1}$] | $\log R'_{\rm HK}$ | Distance [pc] |
|---|---|---|---|---|---|
| 2MASS J00503319+2449009 | 11.22 | 3122.25 | 10.4 | | 14.98 |
| 2MASS J03121265+2951325 | 10.47 | 4201.63 | 18.3 | | 36.55 |
| 2MASS J03413724+5513068 | 10.55 | 4050.0 | 4.5 | | 35.77 |
| 2MASS J05280015+0938382 | 11.18 | 3425.0 | 2.6 | | 10.21 |
| 2MASS J06144242+4727346 | 10.81 | 3739.48 | 7.3 | | 37.3 |
| 2MASS J06192947+1357031 | 10.01 | 3739.48 | 1.0 | | 25.01 |
| 2MASS J08115757+0846220 | 11.38 | 3122.25 | 2.5 | | 6.77 |
| 2MASS J09304457+0019214 | 10.49 | 3275.05 | 1.6 | | 9.91 |
| 2MASS J11285624+1010395 | 11.34 | 3122.25 | 1.7 | | 12.77 |
| 2MASS J11421839+2301365 | 10.83 | 3739.48 | 1.0 | | 30.86 |
| 2MASS J11474440+0048164 | 9.59 | 3122.25 | 3.7 | -5.5 | 3.37 |
| 2MASS J13314666+2916368 | 10.61 | 3122.25 | 55.8 | -4.0 | |
| 2MASS J14511044+3106406 | 11.3 | 3122.25 | 2.2 | | 13.04 |
| 2MASS J18361922+1336261 | 11.15 | 3122.25 | 1.6 | -4.76 | 12.02 |
| 2MASS J20103444+0632140 | 10.92 | 3122.25 | 1.0 | | 16.08 |
| 2MASS J21462206+3813047 | 10.82 | 2971.27 | 1.4 | | 7.04 |
| 2MASS J22232904+3227334 | 10.35 | 3350.0 | 8.5 | -4.1 | 15.22 |
| 2MASS J22561349+5919087 | 10.65 | 3739.48 | 1.0 | | 46.21 |
| 2MASS J23415498+4410407 | 10.37 | 2971.27 | 2.5 | | 3.16 |
| 2MASS J23430628+3632132 | 11.15 | 3122.25 | 2.2 | | 8.36 |
| AD Leo | 8.21 | 4363.0 | 3.5 | | 4.96 |
| AU Mic | 7.84 | 3642.0 | 8.5 | -4.11 | 9.71 |
| BD+33 1505 | 9.35 | 3619.0 | 3.7 | | 18.21 |
| BD-02 2198 | 9.12 | 3866.0 | 3.2 | | 14.23 |
| BX Cet | 10.32 | 3275.05 | 3.0 | | 7.22 |
| CE Boo | 9.13 | 3780.0 | 4.3 | | 9.94 |
| EE Leo | 10.28 | 3122.25 | 2.6 | | 6.96 |
| EQ Peg | 9.04 | 3630 | 16.0 | | 6.26 |
| EG Cam | 9.41 | 3739.48 | 2.3 | | 13.48 |
| EV Lac | 9.0 | 3122.25 | 5.1 | | 5.05 |
| G 168-31 | 10.98 | 3429.2 | 1.1 | | 37.19 |
| G 214-14 | 10.38 | 3739.48 | 1.7 | | 23.68 |
| G 234-57 | 10.46 | 3429.2 | 2.0 | | 23.61 |
| G 234-57 | 10.46 | 3429.2 | 2.0 | | 24.31 |
| G 32-5 | 11.4 | 3122.25 | 5.5 | | 12.16 |
| G 99-49 | 9.9 | 3275.05 | 5.7 | | 5.21 |
| GJ 1 | 7.68 | 3429.2 | 2.8 | | 4.35 |
| GJ 1074 | 10.15 | 3584.18 | 4.0 | | 21.04 |
| GJ 1105 | 10.67 | 3275.05 | 1.9 | | 8.85 |
| GJ 15 A | 7.22 | 3605.5 | 3.7 | | 3.56 |
| GJ 160.2 | 9.2 | 4498.0 | 1.0 | | 25.86 |
| GJ 162 | 9.36 | 4201.63 | 2.4 | | 13.92 |
| GJ 176 | 9.0 | 3679.0 | 12.6 | | 9.48 |
| GJ 180 | 9.93 | 3275.05 | 1.7 | | 11.94 |
| GJ 184 | 9.21 | 3739.48 | 3.5 | | 13.85 |
| GJ 2 | 9.08 | 3875.0 | 1.8 | | 11.5 |
| GJ 205 | 7.1 | 3731.2 | 3.3 | | 5.7 |
| GJ 2066 | 9.12 | 3429.2 | 1.9 | | 8.94 |
| GJ 229 | 7.31 | 3814.0 | 3.1 | | 5.76 |
| GJ 26 | 10.05 | 3429.2 | 2.2 | | 12.67 |
| GJ 273 | 8.59 | 3275.05 | 2.2 | | 3.79 |
| GJ 3138 | 10.19 | 3894.54 | 1.0 | | 28.45 |
| GJ 317 | 10.75 | 3275.05 | 2.8 | | 15.17 |
| GJ 328 | 9.29 | 3739.48 | 3.4 | | 20.5 |
| GJ 3304 | 12.12 | 3122.25 | 24.0 | | 13.1 |
| GJ 3323 | 10.65 | 3122.25 | 2.3 | | 5.37 |
| GJ 358 | 9.63 | 3275.05 | 1.6 | | 9.6 |

Table A.1: continued.

| Name | GAIA $G$ mag | $T_{\rm eff}$ | $\langle v \sin i \rangle$ | $\langle \log R'_{\rm HK} \rangle$ | Distance [pc] |
|---|---|---|---|---|---|
| GJ 3649 | 9.88 | 3584.18 | 1.9 | | 16.68 |
| GJ 382 | 8.33 | 3429.2 | 2.2 | | 7.7 |
| GJ 3822 | 9.83 | 3584.18 | 3.5 | | 20.34 |
| GJ 399 | 10.26 | 3429.2 | 1.7 | | 15.57 |
| GJ 3997 | 9.64 | 3739.48 | 2.7 | | 13.68 |
| GJ 408 | 8.97 | 3122.25 | 2.1 | | 6.75 |
| GJ 4092 | 10.12 | 3739.48 | 2.7 | -4.79 | 28.27 |
| GJ 422 | 10.48 | 3275.05 | 1.2 | -5.83 | 12.67 |
| GJ 433 | 8.89 | 3616.0 | 1.3 | -5.22 | 9.07 |
| GJ 436 | 9.57 | 3416.0 | 1.7 | | 9.77 |
| GJ 450 | 8.85 | 3584.18 | 5.8 | | 8.76 |
| GJ 47 | 9.84 | 4104.0 | 2.0 | | 10.53 |
| GJ 49 | 8.66 | 4055.5 | 2.9 | | 9.86 |
| GJ 494 | 8.91 | 3899.5 | 9.1 | -4.0 | 11.5 |
| GJ 514 | 8.21 | 3727.0 | 1.9 | -5.1 | 7.63 |
| GJ 521 | 9.4 | 3584.18 | 2.9 | | 13.36 |
| GJ 526 | 7.61 | 3634.0 | 2.4 | -5.27 | 5.43 |
| GJ 536 | 8.86 | 4067.0 | 1.7 | | 10.42 |
| GJ 552 | 9.72 | 3429.2 | 2.6 | -5.13 | 14.21 |
| GJ 581 | 9.41 | 3442.0 | 1.8 | -5.77 | 6.3 |
| GJ 588 | 8.27 | 3429.2 | 1.8 | -5.38 | 5.92 |
| GJ 606 | 9.59 | 3584.18 | 2.0 | -4.78 | 13.27 |
| GJ 609 | 11.16 | 3122.25 | 2.0 | -5.76 | 10.02 |
| GJ 628 | 8.79 | 3570.0 | 1.5 | -5.52 | 4.31 |
| GJ 649 | 8.82 | 3696.33 | 2.1 | -5.0 | 10.39 |
| GJ 65 | 10.81 | 2971.27 | 26.4 | | 2.72 |
| GJ 667 C | 9.39 | 4880.0 | 1.4 | | 7.24 |
| GJ 674 | 8.33 | 3275.05 | 1.8 | -5.08 | 4.55 |
| GJ 676 A | 8.87 | 3739.48 | 2.6 | -4.76 | 15.97 |
| GJ 686 | 8.74 | 3584.18 | 2.9 | -5.18 | 8.16 |
| GJ 699 | 8.2 | 3244.67 | 2.5 | -5.56 | 1.83 |
| GJ 70 | 9.9 | 3429.2 | 2.0 | | 11.32 |
| GJ 701 | 8.52 | 3630.0 | 1.9 | -5.09 | 7.74 |
| GJ 731 | 9.38 | 3739.48 | 2.7 | | 15.2 |
| GJ 740 | 8.46 | 3584.18 | 2.3 | -4.63 | 11.1 |
| GJ 752 A | 8.1 | 3275.05 | 2.7 | -5.28 | 5.91 |
| GJ 83.1 | 10.67 | 3122.25 | 2.6 | -4.79 | 4.47 |
| GJ 832 | 7.74 | 3707.0 | 2.0 | -5.07 | 4.97 |
| GJ 846 | 8.4 | 3580.0 | 3.1 | | 10.57 |
| GJ 849 | 9.22 | 3275.05 | 1.7 | -5.4 | 8.81 |
| GJ 876 | 8.88 | 3532.0 | 2.5 | -5.7 | 4.67 |
| GJ 880 | 7.79 | 3750.0 | 2.4 | | 6.87 |
| GJ 908 | 8.15 | 3646.0 | 2.6 | | 5.91 |
| GJ 9122 B | 9.92 | 3739.48 | 3.6 | | 17.22 |
| GJ 9404 | 9.87 | 3739.48 | 2.6 | | 23.84 |
| GJ 9440 | 9.71 | 3429.2 | 2.6 | -5.11 | 16.85 |
| GJ 9793 | 10.04 | 3739.48 | 1.0 | | 30.91 |
| Gl 799B | 9.6 | 3123.0 | 10.2 | | 9.8 |
| Gl 841 A | 9.41 | 3429.2 | 4.2 | | 14.86 |
| HD 154363B | 9.17 | 3584.18 | 2.7 | -5.57 | 10.46 |
| HD 233153 | 8.91 | 5125.96 | 2.7 | | 12.27 |
| HD 265866 | 8.86 | 3275.05 | 1.7 | | 5.58 |
| HD 50281B | 9.09 | 4763.86 | 3.9 | | 8.75 |
| HD 95735 | 6.6 | 3563.5 | 7.3 | | 2.55 |
| HD 98712A | 8.75 | 4875.0 | 4.2 | | 13.72 |
| HIP 12961 | 9.6 | 4131.0 | 1.2 | | 23.41 |
| HIP 57050 | 10.58 | 3122.25 | 1.8 | | 11.02 |
| HIP 79431 | 10.24 | 3275.05 | 1.0 | | 14.56 |
| LHS 3432 | 9.8 | 3429.2 | 4.3 | -4.93 | 8.83 |
| LP 609-71 | 9.61 | 3429.2 | 2.7 | | |
| LP 672-42 | 10.81 | 3275.05 | 1.5 | | 13.47 |

Table A.1: continued.

| Name | GAIA $G$ mag | $T_{\text{eff}}$ | $\langle v \sin i \rangle$ | $\langle \log R'_{\text{HK}} \rangle$ | Distance [pc] |
|---|---|---|---|---|---|
| LP 687-17 | 11.38 | 3122.25 | 2.4 | | 19.57 |
| MCC 549 | 10.28 | 3739.48 | 19.1 | | 38.79 |
| Proxima Centauri | 8.95 | 2990.5 | 2.6 | | 1.3 |
| Ross 45A | 11.29 | 3275.05 | 3.7 | | 22.12 |
| Ross 733 | 10.37 | 3122.25 | 14.0 | | 18.11 |
| TYC 1313-1482-1 | 10.27 | 3739.48 | 1.0 | | |
| TYC 4902-210-1 | 10.01 | 3739.48 | 1.6 | | 31.11 |
| V1054 Oph | 8.27 | 3200 | 2.1 | | |
| V1352 Ori | 10.1 | 3122.25 | 4.7 | | 5.79 |
| VV Lyn | 9.59 | 3429.2 | 4.6 | | 11.99 |
| Wolf 906 | 10.17 | 3429.2 | 1.7 | | 14.48 |
| YZ Ceti | 10.43 | 3122.25 | 2.2 | | 3.72 |

| PI | Programmes |
|---|---|
| Aloisi, Robert | G05115 |
| Barnes, Sydney | G03182 |
| Burt, Jennifer | G03272, G04191 |
| Cloutier, Ryan | G03274, G04214. G05152 |
| Davenport, James | G03227, G04039 |
| Gillen, Edward | G05106 |
| Guenther, Maximilian | G05143 |
| Hambleton, Kelly | G03252 |
| Hermes, James | G04137, G03124, G05081 |
| Holberg, Jay | G03178 |
| Hord, Benjamin | G05015 |
| Howard, Ward | G03174, G04132, G05064 |
| Huber, Daniel | G04103, G03251, G05144 |
| Inglis, Andrew | G04186 |
| Jackman, James | G04142, G04139, G05112, G05114, G05126 |
| Kaltenegger, Lisa | G04147 |
| Kane, Stephen | G03106, G04098 |
| Kiman, Rocio | G05123 |
| Kunimoto, Michelle | G04036 |
| Llama, Joe | G03063 |
| Lopez, Eric | G03126 |
| Macgregor, Meredith | G05070 |
| Marocco, Federico | G04211, G05109 |
| Martin, David | G05071 |
| Mayo, Andrew | G03278, G04242 |
| Million, Chase | G03228 |
| Monsue, Teresa | G04222, G03205 |
| Newton, Elisabeth | G03141 |
| Paudel, Rishi | G04212, G03202 |
| Pepper, Joshua | G04178 |
| Pietras, Malgorzata | G05145 |
| Pineda, J. Sebastian | G03225 |
| Plavchan, Peter | G03263 |
| Prsa, Andrej | G05003 |
| Ramsay, Gavin | G04006 |
| Robertson, Paul | G04059, G04148 |
| Silverstein, Michele | G03226, G04188 |
| Taylor, Jake | G03276 |
| Tovar Mendoza, Guadalupe | G05121 |
| Vanderburg, Andrew | G04200, G03207, G05084 |
| Vega, Laura | G03273 |
| Winters, Jennifer | G04033, G03250, G05087 |

Table A.2: PIs and TESS programmes for which 20 s light curves of our targets were obtained.

# Appendix B: Quasi-periodic pulsation additional candidates

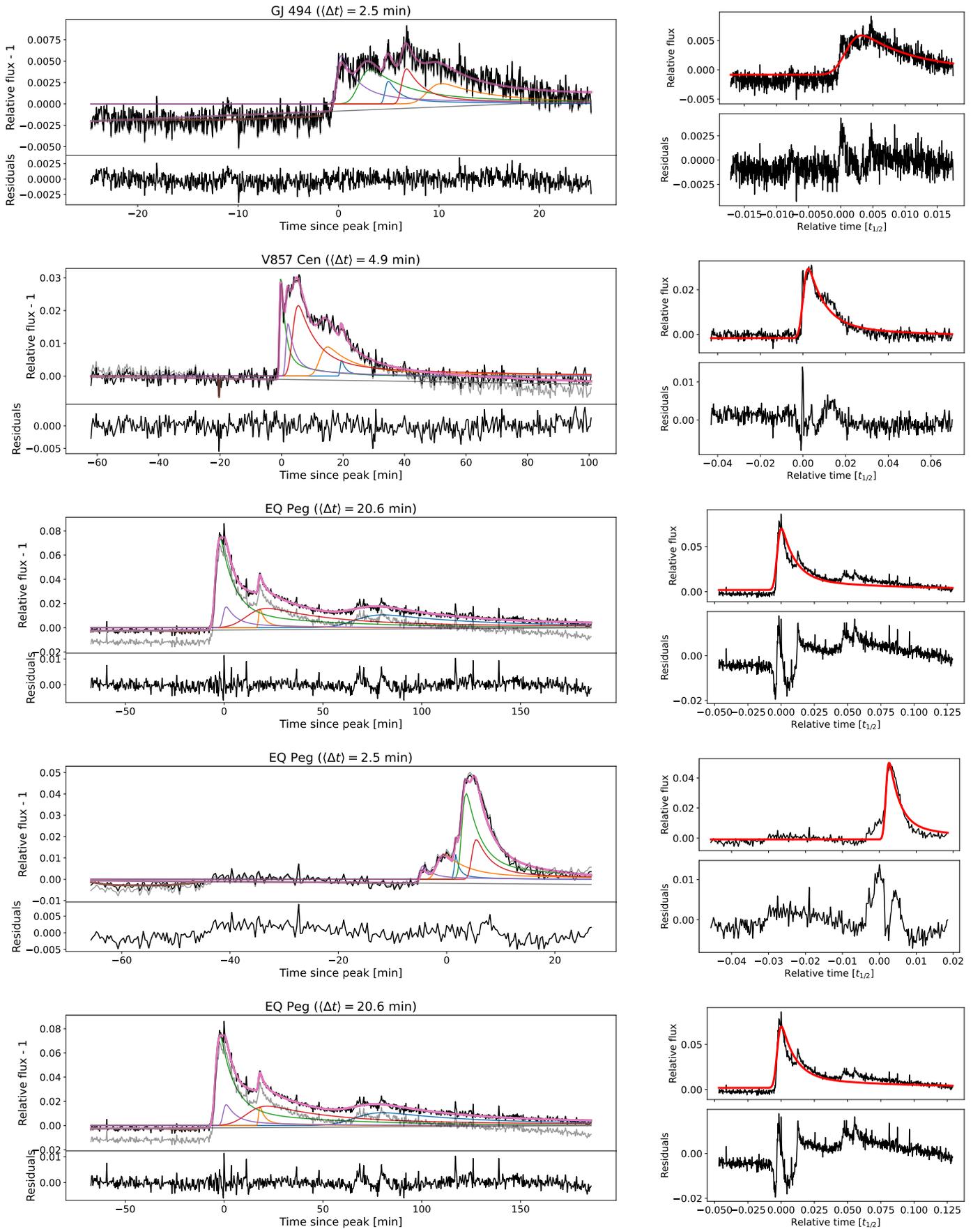

Fig. B.1: QPP candidates not shown in the main text. Each row represents a different candidate, and its description is the same as in Figure 19 (first part).

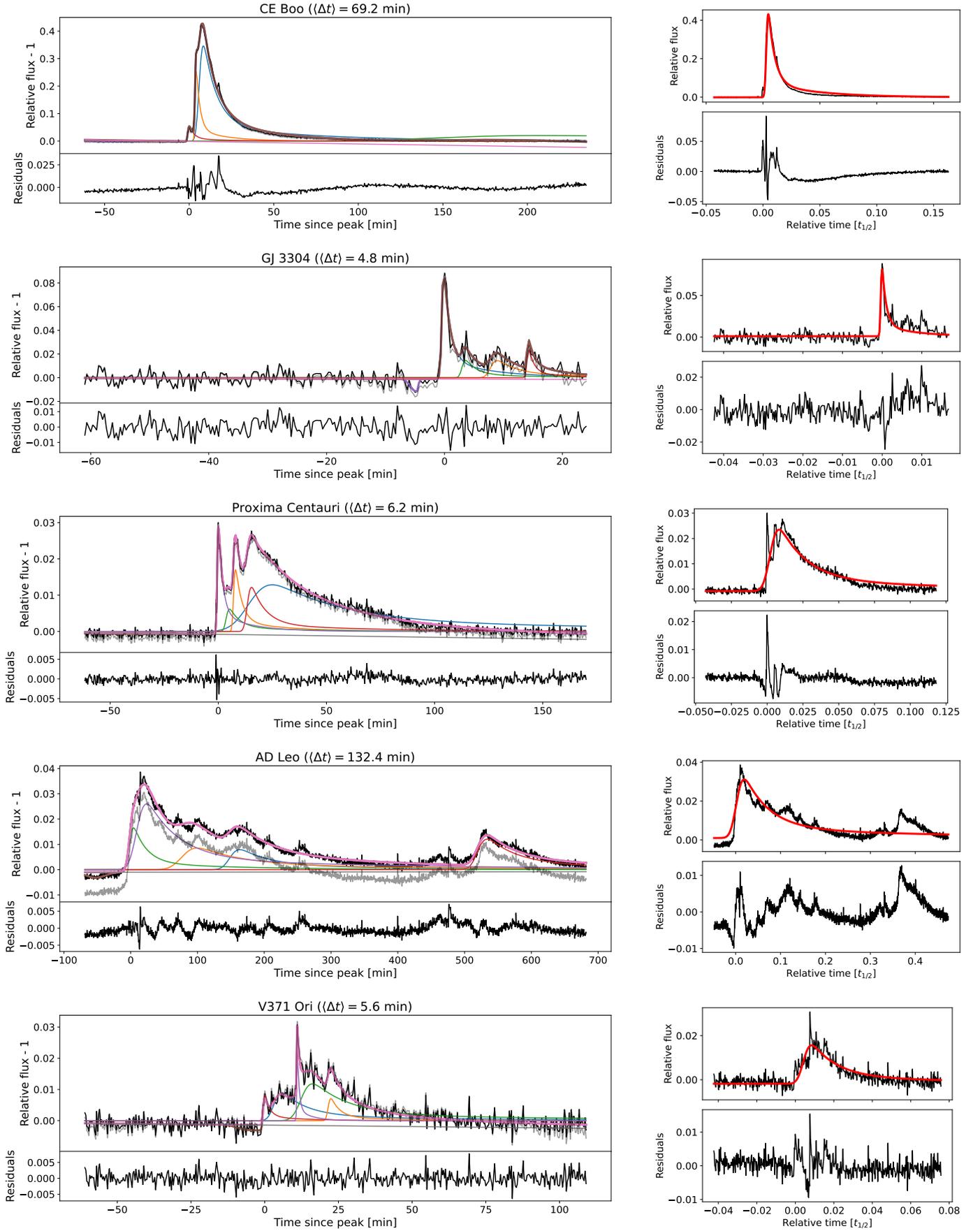

Fig. B.1: continued. The fourth candidate in this Figure (observed on AD Leo) was not used for the calculation of the median flare residual periodicity (Section 6.7), because of its associated poor fit.

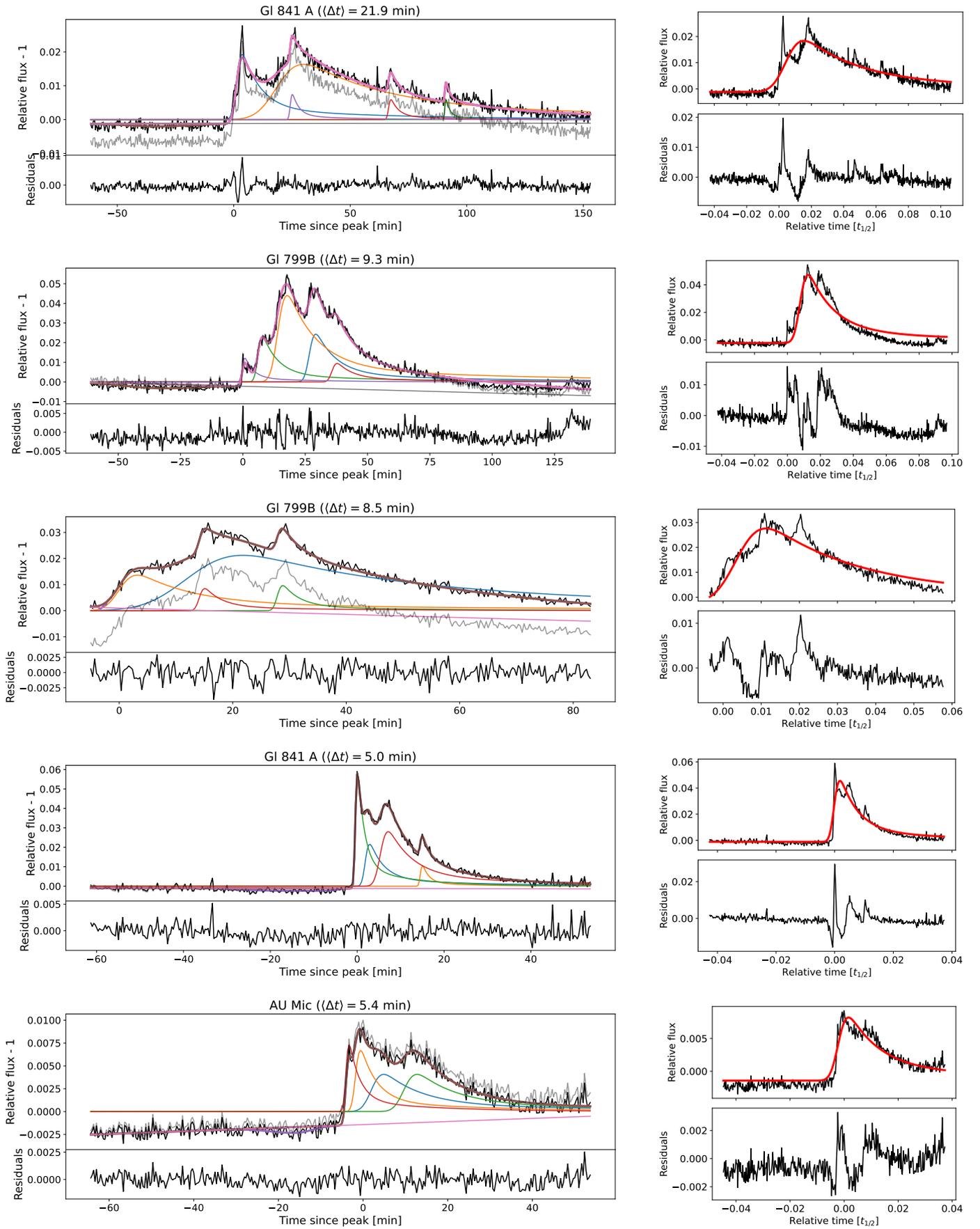

Fig. B.1: continued.